\documentclass[preprint,aps,superscriptaddress,onecolumn,a4paper]{revtex4-1}
\usepackage[colorlinks=true, pdfstartview=FitV, linkcolor=black, citecolor=red, urlcolor=black]{hyperref}
\usepackage{graphicx}
\graphicspath{ {images/} }
\usepackage[all]{xy}
\usepackage{amsmath}
\usepackage{amssymb}
\usepackage{tensor}
\usepackage{orcidlink}
\usepackage{amsthm}
\usepackage{comment}
\newcommand{\cmmnt}[1]{}
\newcommand{\be}{\begin{equation}}
\newcommand{\ee}{\end{equation}}
\newcommand{\ben}{\begin{eqnarray}}
\newcommand{\een}{\end{eqnarray}}
\newcommand{\bes}{\begin{subequations}}
\newcommand{\ees}{\end{subequations}}
\def\bal#1\eal{\begin{align}#1\end{align}}

\newcommand{\bfi}{\begin{figure}}
\newcommand{\efi}{\end{figure}}
\newcommand{\bc}{\begin{center}}
\newcommand{\ec}{\end{center}}

\newcommand{\sech}{{\rm sech}}

\newcommand{\LL}{{\cal L}}

\newcommand{\pu}{\mathrm{\partial_{\mu}}}

\newcommand{\0}{\ensuremath{{\scalebox{0.6}{0}}}}
\DeclareMathOperator{\csch}{csch}

\DeclareMathOperator{\erf}{erf} 
\usepackage{relsize}
\newcommand{\Sj}{\ensuremath{S^{\mathsmaller{\perp}}_j}}
\newcommand{\Si}{\ensuremath{S^{\mathsmaller{\perp}}_i}}
\newcommand{\Sk}{\ensuremath{S^{\mathsmaller{\perp}}_k}}
\newcommand{\pd}[2]{\ensuremath{\frac{\partial#1}{\partial#2}}}

\newcommand{\pb}[1]{\ensuremath{\partial_{#1}}}

\newcommand{\qt}[1]{``#1''}

\newcommand{\Tr}[1]{\ensuremath{\mathrm{Tr}\left(#1\right)}}
\usepackage{calrsfs}
\DeclareMathAlphabet{\pazocal}{OMS}{zplm}{m}{n}

\DeclareMathAlphabet{\mathcal}{OMS}{cmsy}{m}{n}

\newcommand{\g}{\ensuremath{\mathfrak{g}}}
\newcommand{\lm}{\ensuremath{\Lambda_{\mu}}}
\newcommand{\lnu}{\ensuremath{\Lambda_{\nu}}}
\newcommand{\hf}{\ensuremath{\hat{\phi}}}

\begin{document}
\title{Magnetic monopoles in Yang-Mills-Higgs theory with impurities}
\author{D. Bazeia\,\orcidlink{0000-0003-1335-3705}}
     \email[]{bazeia@fisica.ufpb.br}
    \affiliation{Departamento de F\'\i sica, Universidade Federal da Para\'\i ba, 58051-970 Jo\~ao Pessoa, PB, Brazil}
\author{M.A. Liao\,\orcidlink{0000-0001-9720-2079}}\email[]{matheusalvesliao@gmail.com}\affiliation{Departamento de F\'\i sica, Universidade Federal da Para\'\i ba, 58051-970 Jo\~ao Pessoa, PB, Brazil}

\author{M.A. Marques\,\orcidlink{0000-0001-7022-5502}}\email[]{marques@cbiotec.ufpb.br}\affiliation{Departamento de Biotecnologia, Universidade Federal da Para\'iba, 58051-900 Jo\~ao Pessoa, PB, Brazil}

\begin{abstract}
In this work, BPS models built from the coupling of Yang-Mills-Higgs Lagrangian to impurities are investigated. We first consider scalar impurities, which in the BPS limit generate monopoles similar to those obtained in a previously considered class of $\mathrm{SU(2)}\times\mathrm{Z}_2$ or $\mathrm{SU(2)}\times\mathrm{SU(2)}$ models. We then focus on coupling with nonabelian impurities, defined as fixed backgrounds produced from fields transforming under the adjoint representation of SU(2), with a coupling chosen to preserve half of the BPS sectors. The nature of this coupling, the ensuing Bogomol'nyi bound and BPS equations, as well as the effect of these impurities in the abelianization that leads to the emergence of a U(1) gauge group are investigated. We study in greater detail impurities with spherical symmetry, and examine the manner in which impurity coupling changes the asymptotic behavior and range of monopole interactions. Moreover, we introduce a method that can be used to approximate solutions with the use of small perturbations around the Prasad-Sommerfield monopole, and discuss the possibility of extending the aforementioned results to dyons. In order to exemplify the most important properties of the theory, several specific impurity models are presented, with the respective monopole solutions are found numerically. These solutions present novel internal structure and multiple features that would not be possible in the original theory.	
\end{abstract}

\maketitle
\tableofcontents

\newpage
\section{Introduction}
This work deals with magnetic monopoles in Yang-Mills-Higgs (YMH) theory in three-dimensional space in the presence of impurities. Despite their notable absence in standard Maxwell theory, magnetic monopoles have been considered as objects of theoretical interest in physics since the nineteenth century, when their compatibility with electromagnetic theory was first shown by Pierre Curie~\cite{Curie}. They were however mostly viewed as a mathematical curiosity until decades later, when Paul Dirac famously investigated the logical consequences of combining the axioms of quantum theory with the introduction of point-like magnetic monopoles in a system~\cite{Dirac,DiracII}. As Dirac showed, the existence of even one magnetic monopole would suffice to provide a theoretical demonstration of electric charge quantization.
 
Despite Dirac's discoveries, it was not until the 1970s that monopoles were understood as a fundamental part of the spectrum of quantum field Hamiltonians. This perception stems from the realization by 't Hooft~\cite{Hooft} and Polyakov~\cite{polyakov} that monopole solutions appear naturally in Grand Unification Theories (GUTs) as a consequence of the spontaneous symmetry breaking that is fundamental to these models. The need to reconcile the existence of monopole solutions in such theories with the lack of experimental observation of monopoles constitutes the so-called monopole problem. Monopole production in the primordial universe may also lead to important constraints in GUTs~\cite{Einhorn}. The monopole problem was one of the motivations for inflationary models~\cite{Guth}, according to which scattering caused by early inflation may explain the negligible monopole density within the observable universe.

Investigations of non-monopole topological defects in impurity-doped systems have been conducted in a wide range of scenarios. Those include kink solutions emerging within scalar-field configurations~\cite{KinkI, KinkII,KinkIII, AdamI, AdamII, BLMPLB, BLM24}, an idea which the authors have recently extended to two-field configurations~\cite{BLMCSF}. There have also been significant advancements in the understanding of interactions between impurities and the topological vortices that are found in Nielsen-Olesen~\cite{Hook, Tong, Cockburn}, Bose-Einstein~\cite{BEI, BEII} and superfluid~\cite{SCI, SCII} theories. Other topological defects such as Skyrmions~\cite{SkyrmionimpI, SkyrmionimpII}, Instantons~\cite{InstantonI, InstantonII} and sphalerons~\cite{sphaleron} have also been investigated in the presence of impurities.

The case of Yang-Mills monopoles in impurity-doped settings has been investigated in Ref.~\cite{WilsonLines}, where the supersymmetry-preserving addition of Wilson lines to a YMH Lagrangian has been considered. This is done through the inclusion of a term of the form $w^{\dagger}A_{\0}w\delta^3(x)$ in the Lagrangian, where $w$ can be physically interpreted as a spin impurity. This procedure results in the addition of a new source term in the non-Abelian Gauss's Law, thus acting as the YMH equivalent of an impurity. In a similar vein to what the same authors had previously found for vortices coupled to (point-like) electric impurities ~\cite{Tong}, it was then verified that, although the addition of this impurity does not change the Bogomol'nyi-Prasad-Sommerfield (BPS) equations of the theory, it has a significant effect on the system's kinetic energy, with specially interesting applications at low speeds, through the Moduli-space approximation. The magnetic analogue of this theory, achieved through the insertion of 't Hooft lines, can be physically interpreted as Dirac-monopole embeddings, and has also been considered in the past~\cite{hooftlines,  hooftlinesII}.

Although such point-like, singular impurities represented by the insertion of Wilson and 't Hooft line operators have been thoroughly studied in the literature (see, for example,~\cite{WilsonLines, hooftlines,hooftlinesII,hooftlinesIII,hooftlinesIV,hooftlinesV,hooftlinesVI} and references therein), investigations concerning other kinds of impurities appear to be lacking. In this work, we aim to study the effect of nonsingular, localized impurities of nonzero dimension, carrying up an investigation similar to what has been achieved, for example, in the case of kinks~\cite{AdamI}, vortices~\cite{Hook, Tong} and sphalerons~\cite{sphaleron}, where a finite impurity has been introduced to the theory together with an appropriate deformation of the Lagrangian chosen to preserve half the BPS sectors of the model. The ensuing Bogomol'nyi equation is modified by an additive impurity function, which introduces significant qualitative modifications in the internal structure of BPS configurations, and changes the Moduli space generated by these equations. We use a similar framework, resulting in deformed models whose solutions are YMH monopoles in the background generated by adjoint representation of SU(2) impurities.

Other motivations to study magnetic monopoles have appeared in the last two decades. An interesting investigation on the theoretical and experimental status of magnetic monopoles was introduced in Ref. \cite{SI}. In addition, the discovery of localized excitations of the magnetic monopole type in magnetic analog of ice known as spin ice was reported in Refs. {\cite{SIa,SIb,SIc}}. Moreover, in Ref. \cite{SId}, the authors describe the possibility of searching for magnetic monopole production via the Schwinger mechanism. Another experimental possibility is directly related to the Monopole and Exotics Detector at the LHC, the MoEDAL experiment, which was approved in 2010, and the MoEDAL-MAPP detector, with the additional detector MAPP (MoEDAL Apparatus for Penetrating Particles), which was approved at the LHC in 2021 to extend the physics reach of MoEdal by providing sensitivity to millicharged particles and long-lived exotic particles; see, for instance, \cite{MoEa,MoEb} and references therein for more information on this issue. The very recent investigation concerning numerical simulations on collisions of magnetic monopoles, which has appeared in Ref. \cite{CMM}, is also of interest; there, the authors extend the analysis to include the case of relativistic velocities to explore scenarios beyond the BPS limit.

This introduction is followed by a short summary of the standard YMH monopole theory in Sec.~\ref{review}, which will serve a double purpose as a review of the impurity-free theory and a setting in which the conventions here adopted, as well as some important results that will be carried over to the impurity-doped scenario will be introduced. We then introduce the Lagrangian of the general model and discuss the form of the impurity functions in Sec.~\ref{impu}. Two classes of models are considered: those with scalar and adjoint $\rm{SU(2)}$ impurities. Focusing on the latter class, we investigate in the following Sec. \ref{BPSsec} the conditions under which this theory is amenable to a Bogomol'nyi bound, and find that such a bound exists in all multi-monopole (i.e., $N>0$) sectors. The BPS equations solved by the energy minimizers that saturate this bound are found and discussed. In Sec.~\ref{sphere} we introduce spherically symmetric impurities and, using the hedgehog ansatz, find the second and first-order equations in the rotationally symmetric setting. Several properties of these equations were analyzed, including the asymptotic behavior of solutions, which plays a leading role in long-distance interactions involving monopoles. In Sec.~\ref{abel}, we investigate the manner in which the SU(2) symmetry is broken in order to give rise to the U(1) electromagnetic theory outside the monopole core and in Sec.~\ref{ex} the BPS equations are solved for several specific choices of impurities, illustrating some important features of the theory. Moreover, in Sec.~\ref{dyons} we briefly comment on the possibility of describing magnetic monopoles with electric charge. We then finish the work in Sec.~\ref{end} with a brief summary of our findings and perspectives for further research.

\section{Overview of SU(2) monopoles}\label{review}

 We take the well-known YMH theory as our starting point, and work in flat spacetime, with metric tensor $\eta_{\mu\nu}=\mathrm{diag}(-1,1,1,1)$. The scalar field is valued in the Lie algebra $su(2)$ associated to this group, and transforms under the adjoint representation of SU($2$). This means that the group action upon the scalar field corresponds to the mapping 
\begin{equation}\label{gauge}
 	\phi(t,\mathbf{x})\to \g(t,\mathbf{x})\phi(t,\mathbf{x}) \g^{-1}(t,\mathbf{x}), 
 \end{equation}
 where $\g(t,\mathbf{x})$ denotes a gauge transformation. For a complete explanation of Lie theory and the adjoint action, the reader is recommended to~\cite{Lie}.

 In physics literature, $su(2)$ generators are often taken as $t^a=i\sigma_a$, where $\sigma_a$ are the usual Pauli matrices. We shall, however, follow the convention adopted, for example, in~\cite{coleman}, thus writing an arbitrary element $X$ of this vector space as $X=X^aT^a$, where $T^a=-i\sigma_a/2$. Here, and throughout the paper, summation over repeated indices is assumed for both  $su(2)$ and Lorentz indices. It is then a straightforward matter to verify that $\Tr{T^aT^b}=-\frac{1}{2}\delta^{ab}$, which we can use to define the inner product
\begin{equation}\label{product}
	\langle X,Y \rangle=-2\Tr{X Y}.
\end{equation}
Using the cyclic property of the trace, it is easily shown that functions of this product are gauge invariant. This observation shall later aid us in the search for acceptable impurity Lagrangians.

As is well known, the usual partial derivative $\pu$ is not covariant under local gauge transformations. This motivates the introduction of a connection $A_{\mu}$, through the use of which we define the covariant derivative $D_{\mu}\phi = \pu\phi + g[A_{\mu},\phi]$,  
where $g$ is a coupling constant. In order for this derivative to transform in the same way as the scalar field, we require
$
A_{\mu}\to \g A_{\mu}\g^{-1} - \pu\g\g^{-1}
$ under a gauge transformation. 
 
The final element of the Lagrangian is the Yang-Mills field-strength tensor. It represents the curvature induced by the aforementioned connection, in terms of which it can be written as $F_{\mu\nu}=\pu A_\nu - \partial_\nu A_\mu + g[A_\mu,A_\nu]$ \cite{manton}, which is also covariant under gauge transformations. The YMH analogues of the electric and magnetic field are given by $E_k=F_{\0k}$ and $B_k=\frac{1}{2}\epsilon_{ijk}F^{ij}$.

The above definitions can be used to construct the Lagrangian density of the (impurity-free) theory, which is 
  \begin{equation}\label{L0}
 	\LL_{\0}=\frac{1}{2}\Tr{F_{\mu\nu}F^{\mu\nu}} + \Tr{D_{\mu}\phi D^{\mu}\phi} -\frac{\lambda}{4}(1-|\phi|^2)^2,
 \end{equation}
 where we have used the standard quartic potential of Yang-Mills theory, with $\lambda$ being a real constant. Variation of the action obtained from~\eqref{L0} leads to the field equations
 \begin{subequations}\label{SOYM}
 	\begin{align}
 		D_{\mu} D^{\mu}\phi &=\lambda\phi(|\phi|^2-1) , \label{YMsection}\\
 		D_{\mu}F^{\mu\nu} &=g[\phi, D^{\nu}\phi].         \label{YMconn}
 	\end{align}
 \end{subequations}

In order to allow for finite energy configurations, we must fix the boundary conditions in such a way as to limit solutions of~\eqref{SOYM} to those possessing an energy density that goes to zero asymptotically. This is achieved by requiring that solutions approach the set
\begin{equation}\label{vman}
	\mathcal{V}=\{(\phi, A_{\mu}):|\phi|^2=1, A_{\mu}=-\pb{\mu}\gamma\gamma^{-1}\}
\end{equation}
at spatial infinity. If the mapping implied by the boundary conditions is topologically nontrivial, the system contains magnetically charged defects. To make this statement more precise, we recall the analysis conducted by Manton~\cite{MantonII}, who showed that, in any region such that
 \begin{equation}\label{condEM}
	D_{\mu}\hf=0,
\end{equation}
 where $\hat{\phi}=\phi/|\phi|$, the field-strength tensor factorizes into  \begin{equation}\label{factor}
	F_{\mu\nu}=f_{\mu\nu}\hf,
 \end{equation} with
\begin{equation}\label{fmaxwell}
	f_{\mu\nu}\equiv\pu\lnu - \pb{\nu}\lm + \frac{2}{g}\Tr{[\pu\hf,\pb{\nu}\hf]\hf},
\end{equation}
where we have the introduced the functions $\Lambda_{\mu}\equiv\langle A_{\mu}, \hf\rangle$.  The above tensor can be identified with the curvature related to the local $U(1)$ symmetry existing outside the monopole core. It can thus be used to define the electromagnetic fields $e_k=f_{\0k}$, $b_k=\frac{1}{2}\epsilon_{ijk}f^{ij}$ unambiguously in any region in which~\eqref{condEM} holds. By integration of the magnetic field thus defined, one may verify the existence of a magnetic charge 
\begin{equation}
q_m=-\frac{4\pi N}{g}
\end{equation}
associated with any topologically nontrivial solution of finite energy. The number $N$ appearing in this equation is an integer corresponding to the topological charge of a given configuration, and is mathematically understood as the solution's winding number, or topological index~\cite{manton}.

\section{Impurity coupling}
\label{impu}

Let us now extend the results and conventions established in the previous section to the case of a system doped with impurities. From this point onward, we take the coupling constant equal to unity, and note that it can be easily recovered in observable results such as the monopole mass or energy through consistent use of the substitution $F_{\mu\nu}\to F_{\mu\nu}/g$.

We shall consider Lagrangians of the form
\begin{equation}\label{laggen}
\LL=\LL_{\0}+\LL_{imp},
\end{equation}
where $\LL_{\0}$ is as before and $\LL_{imp}$ has explicit dependence not only on the fields, but also on the spatial coordinates, and is meant to model the coupling between impurities and the fields. We also assume that $\LL_{imp}\to 0$ asymptotically, so that interactions with the impurities are localized in a given region of space. Under these conditions, finite energy configurations must still approach elements of~\eqref{vman} asymptotically, so the boundary conditions of the problem, as well as the spontaneous symmetry breaking that gives rise to the U(1) symmetry outside of the monopole core, remain unchanged.

\subsection{Scalar impurities}
We model the impurities appearing in $\LL_{imp}$ through the use of localized, coordinate-dependent functions that increase in those regions in which the system's inhomogeinities are stronger and go to zero away from the location of these impurities, where the system can be considered homogeneous. The simplest way that this can be achieved is through the use of functions that behave as scalars with respect to both Lorentz and gauge transformations. From the discussion in the previous section, we know that the curvature $F_{\mu\nu}$, the scalar field, and its covariant derivative obey a transformation law of the form~\eqref{gauge} when the system is subject to a gauge transformation.  Thus, the simplest gauge-invariant combinations involving scalar impurities are built from simple products between these functions and some quadratic combination of $F_{\mu\nu}$, $D_{\mu}\phi$ and $\phi$. One interesting possibility is 
\begin{equation}\label{impquad}
\LL_{imp}=\text{Tr}\left( S_1(\mathbf{x})B_kB_k +S_2(\mathbf{x})D_k\phi D_k\phi\right),
\end{equation}
where $S_1(\mathbf{x})$ and  $S_2(\mathbf{x})$ are localized scalar functions. In a neighborhood of a given $\mathbf{x}_{\0}$ such that $S_1$ can be approximated by a nonzero constant, the magnetic field defined from~\eqref{fmaxwell} behaves as if in a medium of permeability $\mu^{-1}(\mathbf{x}_{\0})=1 + S_1(\mathbf{x}_{\0})$. Similarly, we can define, as a notational shorthand, a function $\tilde\mu(\mathbf{x})$ such that $\tilde\mu^{-1}(\mathbf{x}_{\0})=1 + S_2(\mathbf{x}_{\0})$. In the static regime, defined by the conditions $E_k=D_{\0}\phi=0$ that follow from the requirement of null kinetic energy~\cite{manton}, the field equations can be concisely written in the form
 \begin{subequations}\label{SOYM2}
	\begin{align}
		D_{k}\left(\frac{D_k\phi}{\tilde\mu(\mathbf{x})} \right)&=\lambda\phi(|\phi|^2-1),\\
		D_{j}\left(\frac{F_{jk}}{\mu(\mathbf x)}\right)&={\frac{\left[\phi, D_{k}\phi\right]}{\tilde\mu(\mathbf x)}},      
	\end{align}
\end{subequations} 
while the $\nu=0$ component of~\eqref{YMconn}, which corresponds to the YM analogue of Gauss's law, is unchanged in both the static and dynamic scenarios. In order for the Lagrangian of the theory to remain well-defined throughout space, we have assumed that $\tilde\mu(\mathbf x)$ is nonzero throughout all space. This assumption also ensures that constant vacuum solutions still exist after the impurities are introduced.

As famously shown by Bogomol'nyi~\cite{bogo}, a special subset of solutions appears in the limit $\lambda\to 0$, which physically corresponds to vanishing Higgs mass. In the present case, a BPS limit also exists if under the same assumptions the model is such that $\tilde\mu(\mathbf{x})=1/\mu(\mathbf{x})$. In that case, one can show, by an almost identical calculation, that the energy functional is subject to the bound
\begin{equation}
	E\geq 4\pi|N|,
\end{equation}
with equality attained if, and only if, configurations are static and satisfy
\begin{equation}\label{eq12}
	B_k \pm \mu(\mathbf{x}) D_k\phi=0,
\end{equation}
with the upper sign giving a $N$ monopole solution (corresponding to positive topological charge $N$), while the lower one results in a configuration with $N$ antimonopoles (defined by negative topological charge).

Equations~\eqref{eq12} are formally contained within a subset of the models encountered in Refs.~\cite{internal, bimag, multimag}, which have been generated through the symmetry enhancement of the SU(2) models investigated in \cite{casana1,casana2,smallandhollow}. In the papers \cite{internal, bimag, multimag}, magnetic monopoles were considered in $\mathrm{SU(2)}\times\mathrm{Z_2}$ or $\mathrm{SU(2)}\times\mathrm{SU(2)}$ theories with generalized derivative terms, and it has been shown that equations identical to~\eqref{eq12} appear in the BPS limit of those theories when the generalized permeability of a given $\mathrm{SU(2)}$ sector is independent of the scalar field of that sector, thus resulting in separable equations. See the aforementioned references for details. Because of this correspondence, we shall not dwell on this system any longer, instead choosing to focus on the SU(2) impurities that will be introduced in the next subsection. That said, it is important to note that these theories do not coincide outside of the BPS regime. 
\subsection{SU(2) impurities}

We now turn our attention to linear impurity couplings, defined by the requirement that such a coupling be expressible in the Lagrangian as a product between impurity functions and the fields. Specifically, $\LL_{imp}$ shall be chosen in such a way as to preserve half of  the BPS sectors of the model. 

Models with this feature have been investigated in Ref.~\cite{Hook}, where impurity-doped magnetic vortices were examined in flat, three-dimensional spacetime. In the aforementioned reference, it was shown that magnetic impurities can be introduced in a way that preserves half of the supercharges of a  $\mathcal{N}=4$ supersymmetric quantum electrodynamical theory that possesses the critically coupled Nielsen-Olesen Lagrangian as its bosonic part. The resulting models were further investigated in Ref.~\cite{Tong}, with emphasis on vortex dynamics. Later, a framework developed along the same lines was introduced in the context of a single scalar field theories~\cite{AdamI}. In both of these models, the impurity coupling results in an asymmetry between configurations possessing positive topological charge (i.e., vortices or kinks and the aforementioned theories) and those of negative charge (i.e., antivortices and antikinks). This asymmetry is reflected in the fact that a Bogomol'nyi bound exists only for configurations whose charge has, say, positive sign, thus explaining the \qt{half-BPS} terminology. 

As discussed in the previous section, scalar impurities are not suitable for use in $\LL_{imp}$ if the couplings are to be linear. We can, however, obtain gauge-invariant expressions through the use of products of the form~\eqref{product}, which can be added in our theory if the impurities are taken as $su(2)$-valued vector fields, which must thus be subject to a transformation law of the form~\eqref{gauge} when acted upon by the gauge group. The simplest such combination that still allows for multi-monopole BPS solutions in the massless Higgs limit is
\begin{equation}\label{Limp}
\LL_{imp}=2\Tr{S_k(\mathbf{x})D_k\phi+S_k(\mathbf{x})B_k},
\end{equation}
where $S_k$, $k=1,2,3$ is a square-integrable impurity function. We also assume that both $S_k$ and its first derivatives are localized in space, thus ensuring that the usual boundary conditions can still be met. The field equations derived from the full Lagrangian density $\LL=\LL_{\0}+\LL_{imp}$ are
 \begin{subequations}\label{EL}
	\begin{align}
		-D_{\0}D_{\0}\phi  + D_{k}\left[ D_{k}\phi+S_k(\mathbf{x})\right]&=\lambda\phi(|\phi|^2-1) , \label{impsection}\\
		D_{\mu}F^{\mu j}-	 \epsilon_{ijk}D_kS_i(\mathbf{x})&=[\phi,D_{j}\phi+S_j(\mathbf{x})], \label{impB}\\
		D_{k}E_k &=[\phi,D_{\0}\phi].         \label{gauss}
	\end{align}
\end{subequations} 
We see that Eq.~\eqref{gauss}, the non-abelian version of the Gauss's law, is not changed after the impurities are added to the system, unlike the remaining Euler-Lagrange equations of the theory.  Since $S_k$ is localized, finite energy solutions are still required to tend to~\eqref{vman} asymptotically, so the topological character of the theory remains largely the same.

An important result is obtained if one takes the covariant divergence of~\eqref{impB}. In the usual Yang-Mills-Higgs theory one can easily show, through the combined use of the remaining equations of motion and Bianchi identities, that this leads to a trivial identity as both sides of the ensuing equation vanish identically. This can be understood as a \qt{covariant conservation law} $D_{\mu}J^{\mu}=0$ for the Yang-Mills current. A similar derivation reveals that an analogous result holds in the presence of the scalar impurities considered in the previous subsection. In the present case, however, two terms survive, leading to a nontrivial relation, namely
\be
\epsilon_{ikj}D_jD_kS_i(\mathbf{x})=\left[B_j, S_j \right].
\ee
Since, by definition, the action of $F_{\mu\nu}$ on an arbitrary su(2) vector is such that $[D_{\mu},D_{\nu}] X=[F_{\mu\nu}, X]$, the above relation can be rewritten in the form
\begin{equation}\label{vinculo}
    [B_j+D_j\phi,S_j]=0.
\end{equation}
Thus, the inclusion of adjoint SU(2) background fields leads to a constraint equation that must be satisfied in order to ensure consistency. This constraint may be understood as a consequence of Noether's theorem: the impurity Lagrangian~\eqref{Limp} is gauge-invariant as long as $S_k$ obey a transformation law of the form~\eqref{gauge}, as was indeed assumed above. For that reason, an infinitesimal gauge transformation implies a variation $\delta\LL=\pd{\LL}{S_{k}}\delta S_k$, which can be shown to vanish identically if~\eqref{vinculo} is satisfied, thus explaining the origin of this constraint.

It should be noted that this equation must be viewed as a constraint on the fields of the theory. Indeed, any acceptable solution must be such that $D_k\phi+B_k$ is parallel to the direction of $S_k$ in $su(2)$. Interestingly, this constraint does not seem to restrict physical solutions at least in the cases we shall deal with, being in fact trivially satisfied by any BPS solution, as can be seen immediately from the first-order equations derived in the next section. This requirement is also automatically satisfied by spherically symmetric monopoles, as will be shown in Sec.~\ref{sphere}.

As discussed in Section~\ref{review}, the unambiguous identification of the electromagnetic field is only possible provided a factorization of the form~\eqref{factor} is achieved far from the monopoles, a property that is derived directly from Eq.~\eqref{condEM}. The full derivation of this property is somewhat tedious, but the important fact is that this results follows directly from the  combination of~\eqref{condEM} with properties of the Lie algebra. Since localized impurities preserve boundary conditions, the relationship between topological and magnetic charges of finite energy configurations remains the same. 

\section{Bogomol'nyi bound}\label{BPSsec}
We now perform a Bogomol'nyi procedure~\cite{bogo} to find the BPS states allowed in the theory with SU(2) impurities. As in the impurity-free scenario, the Bogomol'nyi bound is obtained in the limit $\lambda\to 0$. Since the kinetic energy $T= \frac{1}{2}\int d^3x\left(|D_{\0}\phi|^2 + |E_k|^2\right)$ is nonnegative and can thus only increase the energy, it suffices to minimize the static functional, namely
\begin{equation}\label{E}
	\begin{split}
		{E}=\frac{1}{2}\int d^3x\left(|D_{k}\phi|^2 + |B_k|^2-\Tr{\{S_k,D_k\phi +B_k\}-S_kS_k} \right),
	\end{split}
\end{equation}
where the last term, which evidently does not affect the field equations of the theory, was added to fix the vacuum and ensure that the Bogomol'nyi energy levels, would otherwise be shifted by $\frac{1}{2}\int d^3x|S_k|^2$, are the same for all $S_k$. In the above equation, we have used the cyclic property of the trace to write the $\LL_{imp}$ terms as an anticommutator. Since the square of matrices satisfies the property $(X+Y)^2=X^2+\{X,Y\}+Y^2$, we can combine the last integral in~\eqref{E} to the usual $(D_k\phi + B_k)^2$ expression of the $S_k=0$ case to write
\begin{equation}\label{EBPS}
		E=-\int d^3x\text{Tr} \left[(B_k + D_k\phi + S_k)^2\right] + 4\pi N \geq 4\pi N,
\end{equation}
where $N\geq0$. Equality is attained if, and only if,
\begin{equation}\label{BPS}
	B_k+D_k\phi=-S_k.
\end{equation}
Solutions of the above equation are minimum energy $N$ monopole configurations, with the lack of symmetry under conjugation $N\to -N$ signaling the absence of antimonopole BPS sectors. We note that each Bogomol'myi equation is deformed by an additive impurity contribution, much in the same way as what is found for other impurity-doped defects in half-BPS systems. We also remark that $S_k$ disappears upon linearization, so that the parameter counting arguments leading to a $4N$-dimensional Moduli Space~\cite{WeinbergMP} can be applied here in a straightforward way. As remarked above, the constraint~\eqref{vinculo} is automatically satisfied by any solution of~\eqref{BPS}, since every vector commutes with itself.

In contrast to what is found when other types of defects interact with impurities in the half-BPS regime~\cite{Hook, AdamI,sphaleron}, this procedure does not demand a deformation in the self-interaction potential, which we take in the same form as in the original model. This is, of course, a consequence of the fact that the BPS limit of YMH theory is achieved for null potential. This observations is irrelevant to systems with exact Bogomol'nyi saturation, but the BPS equations are useful as approximations for energy minimizers in more complicated systems that do not contain exact BPS solutions. The Bogomol'nyi procedure also provides a good zero-order approximation for monopole masses~\cite{Mass}, and that application may work for a wide range of self-interaction potentials, as long as a sensible approximation $V(|\phi|)\approx 0$ exists under suitable conditions. Thus, this degeneracy inherited from the impurity-free theory is physically remarkable, in the sense that it allows for greater freedom in the couplings between $\phi$ and the impurity.

\section{Spherically Symmetric monopoles}
\label{sphere}

\subsection{Rotationally invariant energy functional and field equations}
In this section, we look for solutions under the assumption of spherical symmetry, the same path that led 't Hooft and Polyakov to the first topological monopole solution. This is done with the introduction of a spherical ansatz. As a first step, we use the fact that a gauge transformation which changes $A_{\0}$ to zero always exists~\cite{coleman}. When the temporal gauge is fixed in that way, the static condition becomes equivalent to the requirement that all fields be time-independent. It can then be shown that, in an appropriately chosen gauge, spherically symmetric solutions are given by 
\begin{subequations}\label{hedgehog}
	\begin{equation}
		\phi^a = \frac {h(r)}{r}x^a,  \label{hedgehogH}
	\end{equation}
	\begin{equation}
		A^a_i = \frac{k(r)-1}{r^2}\epsilon_{abi}x^b.\label{hedgehogA}
	\end{equation}
\end{subequations}
This is the well-known \qt{hedgehog} ansatz used by 't Hooft and Polyakov in the standard YMH theory but the breaking of translational invariance fundamentally changes this scenario. While the above ansatz technically makes sense for any choice of impurity, we can only reasonably expect solutions of this form when the impurity is spherically symmetric about a given point, which we choose as the origin in our analysis. Formally, this means that the impurity must be chosen to ensure that the energy functional of the static theory is radially symmetric in relation to the origin. 

Although the impurity is in general written as a collection of nine functions, the construction of~\eqref{hedgehog} already presupposes a specific gauge fixing, thus constraining the functional form of $S_k$. Together with the requirement of spherical symmetry, which effectively makes the problem one-dimensional, we are left with three degrees of freedom.  For the purpose of seeking spherically symmetric solutions, it suffices to consider impurities which, in the hedgehog gauge, are written in the form
\begin{equation}\label{symImp}
	S_k=\alpha(r)x_kx^a T^a + \beta(r)T^k,
\end{equation}
where $\alpha(r)$ and $\beta(r)$ are differentiable functions such that $S_k$ remains  regular throughout all space, and which go to zero asymptotically. These impurities  are invariant with respect to rotations about a specific point at the center, which we have used to define the origin of the system of coordinates considered in our calculations. One may inquire about the inclusion of a third arbitrary function of $r$ in a direction orthogonal to~\eqref{symImp}, but it can be verified through direct analysis of the products $\langle S_k, D_{k}\phi \rangle$ and $\langle S_k, B_{k} \rangle$ that the contribution of an eventual third component orthogonal to the above impurity would amount to a physically irrelevant surface term in the action, which could provide no contribution to the spherically symmetric equations. Moreover, explicit calculation of $B_k$ and $D_k\phi$ in the hedgehog gauge shows that both of these quantities are parallel to the direction defined by~\eqref{symImp}, so 
that~\eqref{vinculo} is automatically satisfied. We therefore see that, at least in the symmetric case, this constraint does not restrict the possible solutions even outside of the BPS regime.

Since any impurity defined according to~\eqref{symImp} is localized by hypothesis, and since the monopole is centered at the origin, acceptable configurations must satisfy the boundary conditions
\begin{subequations}\label{BC}
\begin{align}
h(0) = 0, && 	\lim_{r\to\infty}h(r)  = 1, \label{BCa} \\
	k(0) = 1,	 	 && \lim_{r\to\infty}k(r)= 0. \label{BCb}
\end{align}
\end{subequations}

The resulting spherically symmetric energy functional can be written in the form	
\begin{equation}\label{SymE}
	\begin{aligned}
	&E=4\pi\int_{0}^{\infty}r^2dr \Bigg[\left(\frac{k'}{r}\right)^2  +\frac{(k^2-1)^2}{2r^4} +  \left(\frac{kh}{r}\right)^2 +\frac{h'^2}{2} + \frac{\lambda}{4}(h^2-1)^2\\
		& \hspace{90pt}+\frac{2\beta}{r}\left(k'+hk\right) + \left(\alpha r +\frac{\beta}{r} \right)\left(rh'+\frac{k^2-1}{r}\right) \Bigg],
	\end{aligned}
\end{equation}
where the prime denotes differentiation with respect to $r$ and the explicit radial dependence of the functions $\alpha(r)$, $\beta(r)$, $h(r)$ and $k(r)$ has been omitted for simplicity. Since the integrand is a function of $r$ alone, this functional is manifestly symmetric under arbitrary rigid rotations in space. Variation of $E[h,k]$ with respect of the fields gives
\begin{equation}\label{S0h}
	\frac{1}{r^2}\left(r^2 h'\right)' - \frac{2hk^2}{r^2}=\lambda h(h^2-1)+\frac{2\beta}{r}(k-1)-4r\alpha -r^2\pb{r}\alpha -\pb{r}\beta
\end{equation}
and 
\begin{equation}\label{SOk}
	k''+\beta'=kh^2 - \frac{k(1-k^2)}{r^2} +\alpha k r + \beta(k-1),
\end{equation}
which are the spherically symmetric restriction of~\eqref{EL}. In the  $\lambda\to 0$ limit, the above equations are solved by~\eqref{BPS}, which under our assumptions reduce to
\begin{subequations}\label{FO}
	\begin{align}
		h'&=\frac{1-k^2}{r^2} -\alpha r^2 -\beta, \label{FOA}\\
		\frac{k'}{r}&=-\left(\frac{hk}{r}+\beta\right). \label{FOB}
	\end{align}
\end{subequations}
Substitution of these equations into~\eqref{SymE} gives an energy $4\pi$, consistent with saturation of the Bogomol'nyi bound of the $N=1$ sector.

\subsection{ODE reduction and special cases}\label{reduc}
The $S_k=0$ version of Eqs.~\eqref{FO} is satisfied by the celebrated Prasad-Sommerfield (PS) solution~\cite{ps} $(h(r),k(r))=(\coth(r)-1/r,r\csch(r))$. Besides its historical role as perhaps the first nontrivial saturation of a Bogomol'nyi bound, this solution is notable for providing  one of the relatively few examples of a system of nonlinear differential equations with closed-form solution. This cannot be achieved for arbitrary $S_k$, although it is not difficult to find closed-form solutions for conveniently chosen impurities. Nevertheless, most solutions of~\eqref{FO} shall require numerical methods, which will be used in the examples considered in Section~\ref{ex}. However, we now show that in some cases the analysis can be simplified, as the form of these equations allows for reduction of the problem to a single integro-differential equation. To achieve this, we first note that Eq.~\eqref{FOB} can be solved for $h$, leading to
\begin{equation}\label{k}
	k(r)=e^{-\int dr h(r) }\left(1- \int_{r_{\0}}^{r} d\xi \beta(\xi)\xi e^{\int dr h(r)} \right).
\end{equation}
Substitution of this result into~\eqref{FOA} results in a decoupled differential equation of $h(r)$. The problem has thus been effectively reduced to a single integro-differential equation for $h(r)$, since knowledge of this function automatically gives $k(r)$ after integration and substitution into~\eqref{k}. 

Aiming for simplification of Eq.~\eqref{FOA}, we define, without loss of generality,
\begin{equation}\label{htildeDef}
	\tilde{h}(r)\equiv h(r)+\frac{1}{r},
\end{equation}
in terms of which one can write
\begin{equation}\label{htilde}
	\tilde{h}'(r)=-e^{-2\int dr \tilde h(r)}\left(\int_{r_{\0}}^{r} d\xi \beta(\xi) e^{\int dr \tilde h(r)} -1\right)^2  -r^2\alpha(r) -\beta(r),
\end{equation}
which suffices to solve the problem in principle. If $\beta=0$, $k(r)$ simplifies to
\begin{equation}
	k(r)=e^{-\int dr h(r) } \label{k0}
\end{equation}
while~\eqref{htilde} is reduced to the much simpler equation 
\begin{equation}
	\tilde{h}'(r)=-e^{-2\int dr \tilde h(r)} -r^2\alpha(r) \label{h0}.
\end{equation}
In the impurity-free case this equation is reduced to $\tilde{h}'(r)e^{-2\int dr \tilde h(r)}=1$, which is solved by $\tilde h=\coth(r)$ and implies $k(r)=r\csch(r)$, thus recovering the PS solution.

At this point we are still dealing with an integro-differential equation, but it is not a simple matter to transform the above equality into an ordinary differential equation (ODE). To this end, let us define the auxiliary function
\begin{equation}
	y(r)\equiv \int dr \tilde h(r),
\end{equation}
with use of which Eq.~\eqref{htilde} can be mapped into 
\begin{equation}\label{yOde}
		y''(r)=-e^{-2y(r)} -r^2\alpha(r),
\end{equation}
so that we have now reduced the problem to an ODE for $y(r)$, which is simpler to deal with than the original problem of two nonlinear coupled differential equations. If the impurity is sufficiently small, we may treat it as a perturbation affecting the system. This approach allows for use of the PS solution as a zeroth-order approximation supplemented by a small correction. To this end, let $\epsilon<<1$ be a real constant such that 
\begin{equation}
	r^2\alpha (r) = \epsilon \lambda(r) + \mathcal{O} (\epsilon^2)\bar{\lambda}(r),
\end{equation}
where $\lambda(r)$ and $\bar{\lambda}(r)$ must be bounded functions of $r$, which in particular implies that the second term of this equation can be neglected for sufficiently small $\epsilon$. Under this assumption, let us attempt a solution of the approximate form
\begin{equation}\label{correction}
	\tilde h(r) \approx \coth(r) +  u(r),
\end{equation}
where $u(r)$ may be seen as a small correction to the PS solution, representing the perturbation induced by $\epsilon\lambda$. Substitution of the above equation into~\eqref{h0} leads to
\begin{equation}
u'(r) = -\csch^2(r)  \left(e^{-2 \int dr u(r)} - 1\right) - \epsilon \lambda(r).
\end{equation}
Assuming $u(r)$ is small for a sufficiently small $\epsilon$, we can eliminate the exponential through the approximation		
	$e^{-2 \int dr u(r)} \approx 1 - 2\int dr u(r)$, with use of which we can write 
	\begin{equation}\label{37}
		u'(x) = 2\csch^2(r)\int_{r_{\0}}^{r} d\xi u(\xi) - \epsilon \lambda(r),
	\end{equation}
	which is a Volterra integro-differential equation. Equations of this kind have several applications in physics and engineering, and are thus well understood, with many techniques available to aid in their solving and analysis. See~\cite{IntDif} and references therein for more information. 
	
It is now possible to eliminate the integral in~\eqref{37} by means of a strategy similar the one used in the derivation of \eqref{yOde}. We hence define $v(r)\equiv \int dr u(r)$ and write
\begin{equation}\label{40}
	v''(r)= 2\csch^2(r)v(r) - \epsilon \lambda(r).
\end{equation}
Being a linear ordinary differential equation, it can easily be solved for any reasonable choice of $\lambda(r)$, and in many cases it is even possible to obtain closed-form solutions. In general, Eq.~\eqref{40} can be integrated to give
\begin{equation}
		v(r)=\epsilon\left[ I_1(r)\left(1-r\coth(r)\right) + I_2(r)\coth(r)\right],
	\end{equation}
 where $I_1(r)$ and $I_2(r)$ correspond, respectively, to the functions defined by the antiderivatives $I_1(r)=\int dr \lambda(r)\coth(r)$ and $I_2(r)=\int dr \lambda(r)\left(\coth(r)r-1\right)$, so that the solution is completely specified in terms of integrals after $\lambda(r)$ is chosen.
 
\begin{figure}[h]
	\centering
	\includegraphics[width=8cm]{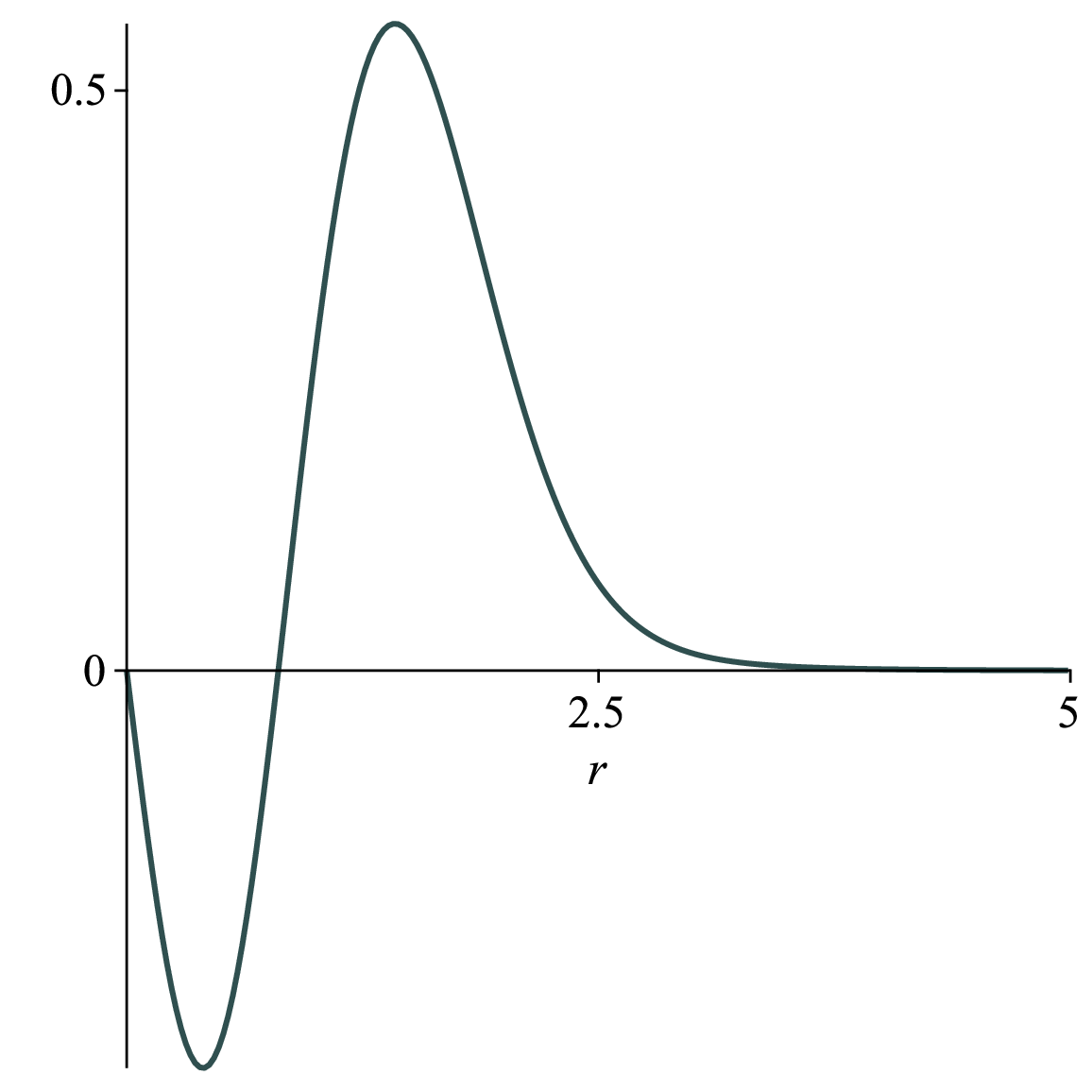}
	\caption{Function $\lambda(r)$, given by~\eqref{lambdafunc}.}
	\label{lambda}
\end{figure}

To test this approximation, let us consider an example that results in a correction $u(r)$ with a simple form. To this end, let us take $\lambda(r)$ in the form
	\begin{equation}\label{lambdafunc}
\lambda(r)= 2r e^{-r^2}\left(r^2 - \frac{\csch^2(r)}{2} - 1\right) + \frac{\sqrt{\pi}}{2}\csch^2(r)\text{erf}(r),
 	\end{equation}
where
\begin{equation}
	\erf(r)=\frac{2}{\sqrt{\pi}}\int_{0}^{r}d\xi e^{-\xi^2} 
\end{equation}
is the error function. The profile of $\lambda(r)$ can be seen in Fig.~\ref{lambda} and asymptotic analysis of~\eqref{lambdafunc} shows that its large $r$ behavior is of the form $\lambda\sim 2\sqrt{\pi}e^{-2r}$, which means that this choice preserves the long-distance behavior of the PS monopole. This function presents a maximum magnitude of $|\lambda(r)|_{\text{max}}\approxeq 0.56$, which leads to $f(r)$ significantly smaller than the boundary values of the fields if $\epsilon$ is of order $10^{-1}$ or smaller, thus being appropriate as a perturbation of the original theory.
Substituting this choice of $\lambda(r)$ into~\eqref{37} and solving for $u(r)$ gives the solution
 \begin{equation}
 	u(r)\approx \epsilon r^2 e^{-r^2}
 \end{equation}
 hence
 \begin{equation}
 	\tilde h(r)\approx \coth(r) +\epsilon r^2 e^{-r^2}.
 \end{equation}

In Sec~\ref{ex}, we have numerically solved a problem with an impurity of the form considered above, for three choices of $\epsilon$, the smallest of which is $\epsilon = 0.1$. As can be seen in Fig.~\ref{ex1}, the approximate result in this case is almost indistinguishable from the numerical plot, while both are similar, but not identical, to the Prasad-Sommerfield solution. The approximation is less precise for bigger values of $\epsilon$, but the other two examples depicted in the same section show that~\eqref{correction} still gives a reasonable result even when $\epsilon=0.5$ or $\epsilon=0.8$ are chosen.

\subsection{Asymptotic analysis}\label{asympt}
In this subsection, we examine the manner in which the addition of SU(2) impurities affects the asymptotic properties of spherical monopole solutions. The long-distance behavior of the fields plays a large role interactions between monopoles, particularly so when their separation is large, which is a very common starting point for dynamical investigations of defects. If the separation is sufficiently large, the force between defects is small enough for an approximation of the system as superpositions of static defects be sensible. Both the characteristic distance above which the separation can be considered \qt{large} and the strength of these interactions at this distance are fundamentally dependent on the asymptotic behavior of static solutions. See, for example,~\cite{MantonII} for an application to YMH monopoles of the method discussed above.

In the scenario we are presently dealing with, impurities also play an important role, as the forces resulting from their coupling are known to produce important effect on defect scattering of all kinds. Spherically symmetric impurities have a greater effect on monopoles located close to the origin, but their range may be sufficiently great to produce long-distance forces as well. If $\alpha$ and $\beta$ fall to zero sufficiently fast (for example, at an exponential rate), the forces generated by these localized objects on a second monopole or antimonopole located far from the origin will be negligible compared to the long-range interactions produced by the Higgs and gauge fields associated with the monopole. In other cases, impurities may fall to zero sufficiently slowly to generate forces with range comparable to that of the fields, and may thus play a significant role in the scattering of all monopoles in the system.

Although the present work is not devoted to dynamics, the above considerations highlight the important role played by asymptotic properties in defect theory, so it shall prove worthwhile to dedicate a few paragraphs to the analysis of these properties. To this end, let $\eta(r)$ and $\kappa(r)$ represent small perturbations such that $h(r)=1+\eta(r)$ and $k(r)=\kappa(r)$ hold to dominant order for large values of $r$. Linearization of Eqs.~\eqref{FO} with these substitutions leads to
\begin{subequations}\label{asymptotic}
	\begin{align}
		\eta'(r)&= \frac{1}{r^2} - f(r) + \mathcal{O}(\kappa^2), \\
		\kappa'(r)&=-\left(\kappa + r\beta(r)\right) +\mathcal{O}(\eta\kappa),  \label{kappap}
	\end{align}
\end{subequations}
where we have defined
\begin{align}
	f(r)\simeq r^2\alpha(r)+ \beta(r) \hspace{20pt} (r>>1),
\end{align}
in which only the dominant contributions to $\alpha$ and $\beta$ at large $r$ need to be considered. Direct integration of these equations lead to 
\begin{subequations}\label{Asympt}
\begin{align}
	\eta(r) &= -\left(\frac{1}{r} +\int_{r_{\0}}^{r} f(\xi) d\xi \right), \label{Asympth}\\
	\kappa(r) &=Ce^{-r} - e^{-r} \int_{r_{\0}}^{r} e^{\xi} \xi\beta(\xi) d\xi, \label{Asymptk}
\end{align}
\end{subequations}
\noindent where $r_{\0}$ is an arbitrary constant. These equations give the lowest-order approximation for the long-distance behavior of the fields. The first term in each of these equations is consistent with the Prasad-Sommerfield solution~\cite{ps}, to which these monopoles must reduce when $f(r)=g(r)=0$. Note that the long-range behavior of $h(r)$ is preserved to dominant order unless the $-1/r$ factor in the BPS equation is canceled.

To exemplify the way in which the long-range behavior of the solution may be changed due to impurity interactions, let us consider functions with a converging power-series representation at infinity, and write
\begin{subequations}\label{exp}
\begin{align}
	f(r)&=\frac{C_m}{r^m}+\frac{C_p}{r^p} +\mathcal{O}\left(\frac{1}{r^{p+1}}\right), \label{alphaexp}\\
	\beta(r)&=\frac{C_n}{r^n}+\mathcal{O}\left(\frac{1}{r^{n+1}}\right), \label{betaexp}
\end{align}
\end{subequations}
\noindent where $p$, $m$ and $n$ are positive integers such that $p>m$, $p,n>1$, and we take the smallest nonzero coefficients in the asymptotic expansion of $\alpha$ and $\beta$, assuming such coefficients exist. If the impurities fall to zero at an exponential or faster rate, then all the above coefficients are zero, and the asymptotic behavior of the monopole is unchanged in relation to the impurity-free theory. Physically, this implies that imourity coupling vanishes outside of the monopole core, where the $\rm{U(1)}$ symmetry is realized (see section~\ref{abel} for details). There are also some noteworthy cases in which expansions of the form~\eqref{exp} may not be valid. Examples include functions with infinitely many zeros at arbitrarily large values of $r$, as is the case of trigonometric integrals and Bessel Functions. In such cases,~\eqref{Asympth} can still be used, but the integrals appearing in these formulas must be treated in a case-by-case basis. Such an analysis is however reasonably simple, because asymptotic expansions of these functions are well-known in the literature (see, for example, Ref.~\cite{NISTI}). Finally, a more complicated situation may be encountered if logarithmic powers or expressions of the form $r^{-a}\ln^{-b}(r)$ are considered. In such cases, the asymptotic analysis is not as straightforward, for such contributions cannot be adequately described in terms of power series. However, it is still relatively simple to deal with impurities of this kind, a simplicity that is largely owed to the simple form of equations~\eqref{FO}. We shall exemplify this situation in the end of this section.


By substituting~\eqref{betaexp} into~\eqref{Asymptk} and solving the integral, we find
\begin{equation}
	k(r)= -C_nr^{-n+1} + \mathcal{O}(r^{-n}),
	\end{equation}
where we have neglected terms of order $r^{-n}$ and greater. We thus see that, if the impurities have a nontrivial power series asymptotic expansion, the range of $k(r)$ is increased, as this function now falls to zero under a power law. Given the association between monopole core size and the asymptotic behavior of $k(r)$ (see next section), we may conclude that the monopole core is enlarged in these cases.

In regard to the long-distance behavior of $h(r)$, a few distinct possibilities exist. First, we note that, if $m>2$, then the integral in~\eqref{Asympth} amounts to contributions of higher order, and thus the asymptotic behavior of $h(r)$ is unchanged by the addition of impurities. If $m=2$, but $C_m\neq 1$, then the scalar field still falls of as $1/r$, albeit with a different coefficient, notably implying  a descending $h(r)$ if $C_m-1>0$. Finally, if $m=2$ and $C_m=1$, then both $1/r$ factors in~\eqref{Asympth} cancel out, resulting in a scalar field that tends to $\mathcal{V}$ at a faster rate. 

Moreover, some impurity functions are such that Eqs.~\eqref{asymptotic} lead to a more accentuated long-distance behavior, extending the range of interactions even further. In these cases, one finds that $\eta$ falls off more slowly than any integer power of $r$, giving rise, for example, to half-integer or even logarithmic tails. In fact, it is easily verified that if one chooses $\alpha$ and $\beta$ in such a way that
\begin{equation}\label{implog}
	f(r)\sim \frac{c}{r\ln(r)^2},
\end{equation} 
 for some real constant $c$, then 
\begin{equation}\label{hlog}
	\eta(r) \sim \frac{c}{\ln(r)}-\frac{1}{r},
\end{equation}
which at very large $r$ is dominated by the second term, resulting in a scalar field that reaches its asymptotic value logarithmically. In Refs.~\cite{Gani,superlong}, other topological structures with logarithmic falloff in scalar field models in $(1,1)$ dimensions were investigated; they were called super long-range kinks. Because $\eta(r)$ falls to zero more slowly than any power of $r$, it may not be reasonable to neglect terms of order $(\kappa\eta)$ in the asymptotic limit, so Eq.~\eqref{kappap} should be modified to 
\begin{equation}
		\kappa'(r)\sim-\kappa(r)\left(1+\eta(r) \right) - r\beta(r),  \label{kappalog}
\end{equation}
which can be completely solved for any well-defined $\eta$ in terms of $\int dr\eta(r)$. In particular, if $\eta(r)$ satisfies~\eqref{hlog} asymptotically, then this equation leads to
\begin{equation}\label{39}
\kappa(r)\sim e^{\text{li}(r)c -r}\left(C_1-\int_{r_{\0}}^{r}d\xi \ e^{\xi-\text{li}(\xi)c}\xi\beta(\xi)\right),
\end{equation} 
where $\text{li}(r)\equiv \int_{\0}^{r}\frac{dt}{\ln(t)}$ is the logarithmic integral~\cite{NISTII}. This expression can be further simplified through use of the relationship between this function and the Exponential Integral, which one may use to write $\text{li}(r)\sim {r}/{\ln(r)}$, although that is generally not sufficient to make the integral in~\eqref{39} solvable in closed-form.

\section{Abelianization}\label{abel}
Let us now examine how, outside the monopole core, the fields abelianize in order to give rise to an effective $\rm{U(1)}$ theory that can be properly identified with electromagnetism. As discussed in Section~\ref{review}, the physical electromagnetic field can be unambiguously defined when the factorization~\eqref{factor} holds. The derivation of this result is somewhat tedious, but it suffices to stress that it follows directly from the combination of~\eqref{condEM} and some basic properties of the lie algebra considered. Since neither of these is affected by the impurity, one can safely conclude that the fields still abelianize in any region such that~\eqref{condEM} is satisfied. By extension, the definition and quantization of magnetic charge, as well at its relationship with the topology of the asymptotic mappings, are unchanged. 

Despite the aforementioned similarities, a more thorough examination of this abelianization process reveals that some important differences may emerge due to impurities, specially in intermediate regions between a given zero of $\phi$, which identifies the center of a monopole, and infinity, where configurations must satisfy $(\phi, A_{\mu})\in\mathcal{V}$. Such an intermediate region would be defined by the requirement that $D_{\mu}\hat{\phi}\approx 0$ to some specified order of approximation, which is not a strong enough condition to trivialize the field equations. In the standard theory, it can be verified that gauge fields fall off exponentially fast, allowing for the identification of a smaller $\rm{U(1)}$ symmetry that survives outside the monopole core. This defines a region in which the usual Maxwell equations are satisfied by $f_{\mu\nu}$, while the Higgs field is governed by a one-component real scalar equation~\cite{manton}. 

In the present scenario, the above reasoning is complicated by the fact that the validity of the relation $D_{\mu}\hat{\phi}\approx 0$ is dependent not only on the specific form of $S_k$, but also on the specific way the zeros of $\phi$ are distributed over space, as these choices are not equivalent anymore. To examine this in more detail, let us consider a region where $\phi$ has no zero and assume $D_{\hf}=0$. It will prove convenient to decompose the impurity functions in the portions parallel and orthogonal to $\hf$, namely,
\begin{equation}\label{impDecomp}
	S_k(\mathbf{x})=s_k(\mathbf{x})\hf + S^{\perp}_k(\mathbf{x}),
\end{equation}
where $s_k$ is a real function while $S^{\perp}_k(\mathbf{x})$ is an $su(2)$ vector such that $\langle S^{\perp}_k, \hf\rangle =0$. Since the terms that make up the impurity contributions to the Lagrangian are built from trace products, the contribution given by $S^{\perp}_k$ vanishes from $\LL$ when~\eqref{condEM} holds. In fact, it is a simple task to verify that $D_{\mu}\hf=0\implies D_{\mu}\phi =\partial_{\mu}|\phi|\hf$. Substituting this result and~\eqref{factor} in~\eqref{Limp} and writing the impurity as~\eqref{impDecomp}, we can derive the effective Lagrangian
\begin{equation}
\LL_{imp}\approx - \left(\pb{k}|\phi|+ b_k\right)s_k,
\end{equation}
which is thus made up of two additive terms, the first of which  consists of the derivative coupling present in half-BPS scalar field theories~\cite{AdamI}, while the second is a magnetic impurity coupling with precisely the same form as that originally derived in the context of half-BPS supersymmetric mirror theories and often considered in connection to vortices~\cite{Hook, Tong}. Thus, as the Abelian symmetry emerges from the condition $D_{\mu}\hf=0$, it becomes possible to identify a corresponding magnetic impurity in the effective large-distance Lagrangian, thus providing an interesting bridge between the models presently considered and the abelian impurity theories.

Next, we examine the form taken by the field equations when~\eqref{condEM} is assumed. It is first worthwhile to note that this condition leads to gauge fields of the form
\begin{equation}
	A_{\mu}=a_{\mu}\hf + \left[\pu\hf, \hf\right]. \label{Amu}
\end{equation}
We can use the above result to derive the covariant derivative of the impurity functions, which are needed in~\eqref{EL}. By expanding $S_k$ according to~\eqref{impDecomp} and making use of $D_{k}\hf =0$, we find
\begin{equation}\label{DS}
	D_k S_j =\left(\pb{k}s_j\right)\hf +\pb{k}\Sj + a_k[\hf,  \Sj ]+[[\pb{k}\hf,\hf],\Sj ].
\end{equation}
By using the Commutator relations to expand out the last term in this equation, we can write
\begin{equation}
	\begin{split}
	[[\pb{k}\hf,\hf],\Sj] &=-2\Tr{\pb{k}\hf \Sj}\hf \\
	&=2\Tr{\hat\phi \pb{k}\Sj}\hf,
	\end{split}
\end{equation}
where in the last step we used the property $\pb{k}\hf^a S^{\mathsmaller{\perp}a}_j=\left(\pb{k}S^{\mathsmaller{\perp}a}_j\right)\hf^a$, which follows directly from differentiation of the relation $\hf^a S^{\mathsmaller{\perp}a}_j=0$. The last line of the above equation is the component of $\pb{k}\Sj$ along the direction defined by $\hf$. Substituting this result into~\eqref{DS} we find, after some manipulations,
\begin{equation}\label{DS2}
	D_k S_j =\left(\pb{k}s_j\right)\hf + a_k[\hf, \Sj]+[\hf,[\pb{k} \Sj,\hf]].
\end{equation}
Substituting~\eqref{condEM} and~\eqref{DS2} into~\eqref{impsection} and projecting the resulting equation into $\hf$ direction, we are led to
\begin{equation}\label{waveEq}
	\Box |\phi| +\lambda |\phi|(1-|\phi|^2)=\pb{k}s_k
\end{equation}
but, unlike what happens in the standard theory, orthogonal directions do not give trivial equations for arbitrarily-chosen impurities. Indeed, the remaining contributions of~\eqref{DS2} give rise to the equations
\begin{equation}\label{constraint}
[a_k\Sk +[\pb{k} \Sk,\hf],\hf]=0
\end{equation} 
which, despite being explicitly dependent on $a_k$ and a partial derivative, is a gauge invariant expression. 

A similar derivation can be applied to~\eqref{impB}. After projecting the resulting equation into the $\hf$ direction, we are led to
\begin{equation}\label{maxwellamp}
\frac{\partial e_j}{\partial t}+\pb{k}\left(f_{k j}-\epsilon_{ijk}s_i\right)=0,
\end{equation}
with the consistency condition
\begin{equation}\label{constraintII}
\left[\epsilon_{ijk}\left(a_k\Si + [\pb{k} \Si,\hf]\right) -|\phi|\Sj,\hf\right]=0.
\end{equation}

We therefore conclude that the field equations can only be consistent with the condition~\eqref{condEM} if the additional equations~\eqref{constraint} and~\eqref{constraintII} are satisfied. Our assumptions ensure that these constraints will always be met at sufficiently long distances, but the distance after which $D_{\mu}\hf\approx 0$ can be used is heavily dependent both on the form of the impurities and their relative distance to the monopole centers. If, however, one considers impurities that fall off sufficiently fast outside of a given neighborhood containing the zeros of $\phi$, then these constraints are trivially solved outside this neighborhood. A particularly interesting class of problems is obtained if compact support is assumed for $S_k$ and its derivatives, which in this case amounts to the requirement that these functions vanish outside a closed, bounded subset of space. Such impurities would of course not contribute to the field equations outside of their support, so that the U(1) symmetry can be identified at a finite distance from the nearest monopole.

In order to shed some light on the above discussion, it is worthwhile to consider the spherically symmetric scenario, in which case this matter is greatly simplified, largely as a consequence of the coincidence  between the centers of $S_k$ and the monopole. In this case, we shall see that equations~\eqref{constraint} and~\eqref{constraintII} are automatically satisfied whenever~\eqref{condEM} is, so that these constraints do not in fact incur any limitation to symmetric models. To proceed, we note that the ansatz~\eqref{hedgehog} implies
\begin{equation}\label{Dkhphi}
	D_{k}\hf =\frac{k(r)}{r}\left(T^k-\frac{x_kx^a}{r^2} T^a\right),
\end{equation}
while the form of~\eqref{symImp} in the hedgehog gauge leads to
\begin{equation}
	\Sk=\beta(r)\left(T^k-\frac{x_kx^a}{r^2}\right)\implies D_k\Sk=-\frac{2\beta(r) k(r)}{r}\hf.
\end{equation}
Thus, it is clear that $\Sk$ and $D_{k}\hf$ disappear from the field equations if and only if both $\beta(r)$ and $k(r)$ vanish in the region of interest. Although these might initially seem like two independent conditions, insertion of $k(r)\approx 0$ into~\eqref{SOk} followed by integration reveals that $k(r)\approx 0 \implies \beta(r)\approx 0$. Thus we see that, in the spherically symmetric case, abelianization occurs if and only if $k(r)\approx 0$ in the hedgehog gauge, the same condition that ensures $D_{\mu}\hf\approx 0$, and thus $F_{\mu\nu}\approx f_{\mu\nu}\hf$, in the standard theory. For a symmetric BPS solution, the monopole core size is entirely specified by~\eqref{Asymptk}. As we see from the form of~\eqref{Asympt}, impurities may modify the asymptotic properties of $k(r)$ to make this function go to zero at a shorter or longer range. Physically, this means that monopole core may be enlarged, shrunk, or even restricted to a finite ball (as happens if $k(r)$ has compact support), depending on the choice of impurities. In the particular cases where impurities fall to zero more slowly than $k(r)$, the exterior region is populated only by a radial magnetic field and a kink-like scalar field obeying~\eqref{waveEq}, with information about the impurities being obtainable from comparison between this scalar field and that of the original YMH theory.

We remark that equations~\eqref{maxwellamp}, taken together, should be physically identified with the Abelian Ampère-Maxwell law of electromagnetism, which is the only among Maxwell's equations to be changed due to interaction with magnetic impurities, which, as we have seen, do not affect Gauss's law. As in the original theory, Faraday's and magnetic Gauss's laws follow directly from Bianchi identities~\cite{MantonII}
\begin{equation}\label{Bianchi}
		\pb{\nu}f_{\mu\lambda} + \pb{\mu}f_{\lambda\nu} + \pb{\lambda}f_{\mu\nu}=0
\end{equation}
which, being a mathematical consequence of the definition of $f_{\mu\nu}$ and its role as a curvature form, are unaffected by the introduction of impurities. Because in particular $\pb{k}b_k=0$, there can be no magnetic sources in any region such that~\eqref{factor} holds, so that all such sources must be contained within the monopole core. A distinction can hence be draw between the models we are working with and those considered in~\cite{hooftlines}, as the latter amount to insertion of charged point-like magnetic sources rather than the currents generated by~\eqref{Limp}. Being geometrical in nature, identities~\eqref{Bianchi} will hold unless the curvature form $f_{\mu\nu}dx^{\mu}\wedge dx^{\nu}$ is ill-defined, as happens in singular points of the theory bundle.  A model governed by a Lagrangian of the form~\eqref{laggen}, where impurities and their derivatives are regular functions, does not give rise to singularities that could be understood as new monopoles, although such a possibility may ensue if the field equations of the theory are generalized to accommodate Dirac-like singularities, which do not allow for a Lagrangian description.

\cmmnt{A similar issue has been considered in Ref.~\cite{Cockburn}, where the authors have investigated the addition of rotationally symmetric delta impurities to an Abelian-Higgs system, obtaining a result that is formally similar to the addition of $\alpha$ fixed vortices whose zeros coincide with the position of the delta impurities. }

\section{Specific models}\label{ex}
We will now solve the first-order equations for a few impurity choices, in order to illustrate some of the theoretical results discussed above in the text. 

\subsection{PS-decaying solutions}
\begin{figure}[h]
	\centering
	\includegraphics[width=0.33\linewidth]{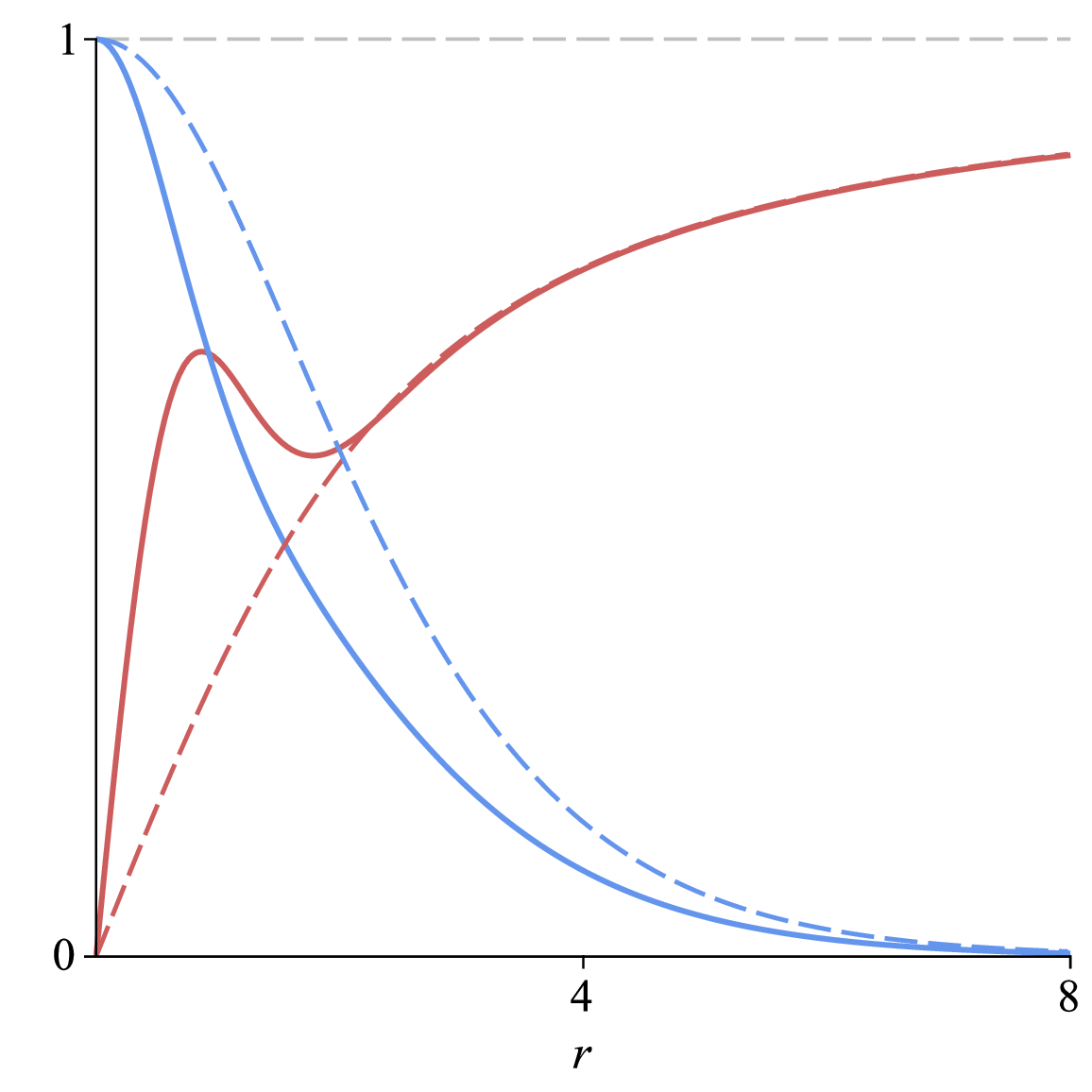}\includegraphics[width=0.33\linewidth]{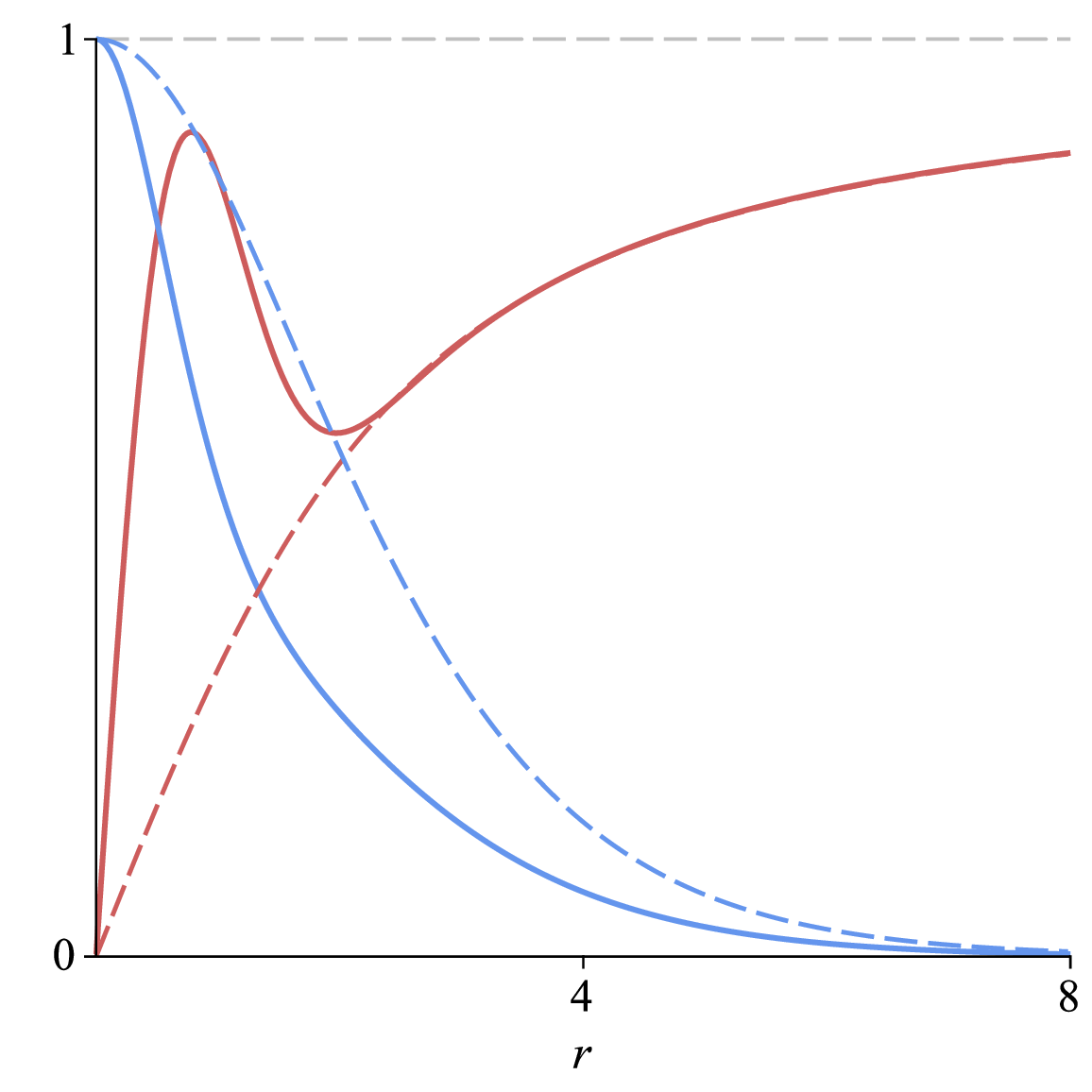}
    \includegraphics[width=0.33\linewidth]{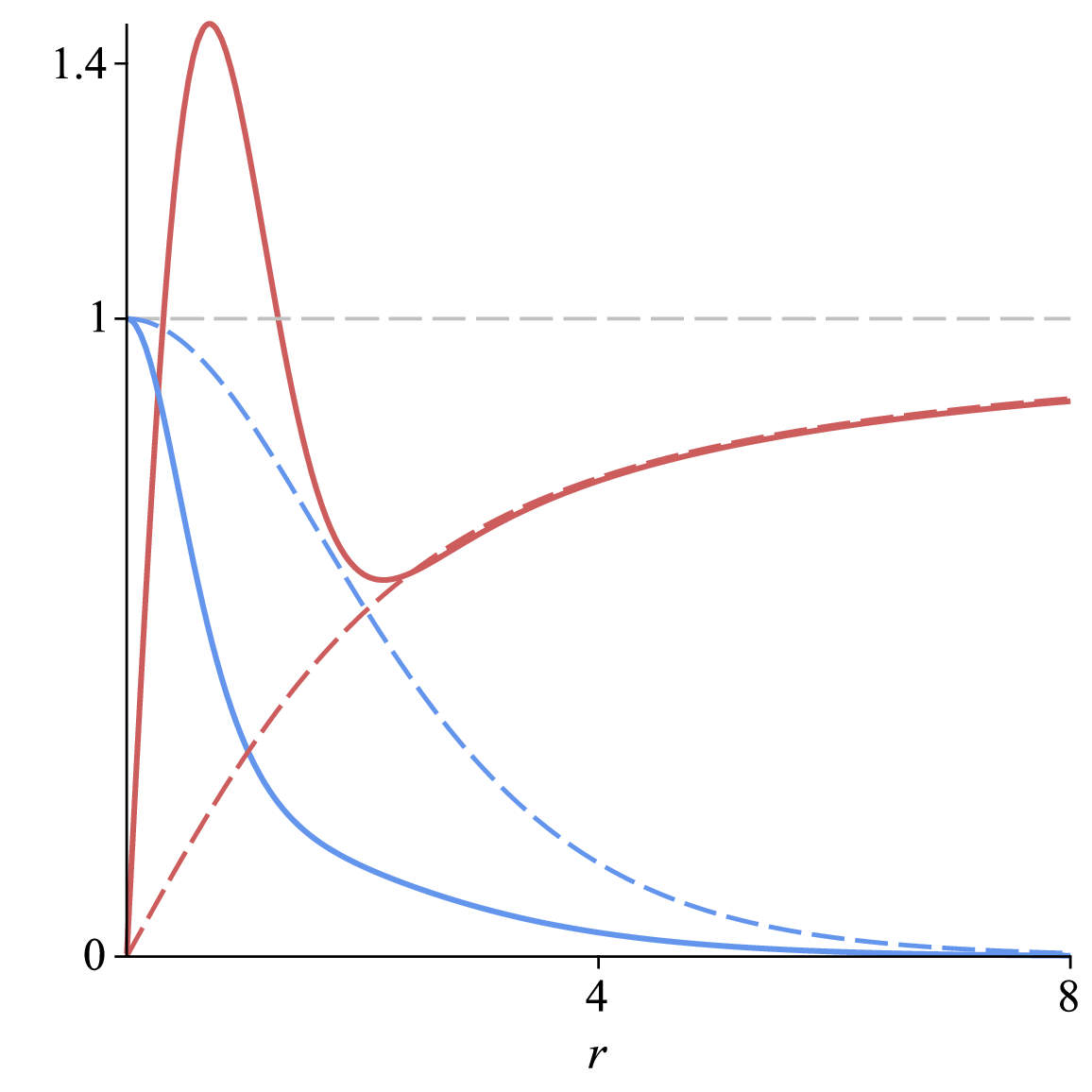}
	\caption{Solution $h(r)$ (red, solid line), $k(r)$ (blue, solid line) of equations~\eqref{FO} with $\alpha(r)$ and $\beta(r)$ given by~\eqref{exponentialImp}, with $A=2$ (left), $3$ (middle) and $4$ (right). Dashed lines of these same colors represent the Prasad-Sommerfield solution.}
	\label{ex1A}
\end{figure}

Let us first deal with impurities that fall to zero exponentially fast, thus giving rise to solutions whose asymptotic behavior is not modified by impurity interaction. Among the simplest examples of this class are the models specified by the choices
\begin{align}\label{exponentialImp}
	\alpha(r)&=Ae^{-r^2},\\ \beta(r)&=0,
\end{align}
where $A$ is a real constant. Due to the exponential decay of $\alpha(r)$, every coefficient in~\eqref{exp} is zero, so that impurity coupling is negligible at large distances, being important only within the core of the monopole. At short distances, however, the field equations are significantly modified, giving rise to a monopole with a novel internal structure. The spherically symmetric BPS solutions have been numerically solved for $A=2$, $A=3$ and $A=5$, and the respective profiles of these solutions are displayed in Fig.~\ref{ex1A}. We see that new maxima and minima emerge inside a neighborhood of the impurities, a behavior that is notably different from the monotonic character of the PS solution, $(h(r),k(r))=(\coth(r)-1/r,r\,{\rm csch}(r))$. It is also notable that the peak of $h(r)$ increases with different choices of $A$, getting higher as $A$ is increased. We have tested this trend for many values of $A$, although only three have been depicted here, in the interest of cleanness of presentation. 

\begin{figure}[h]
	\centering
	\includegraphics[width=0.33\linewidth]{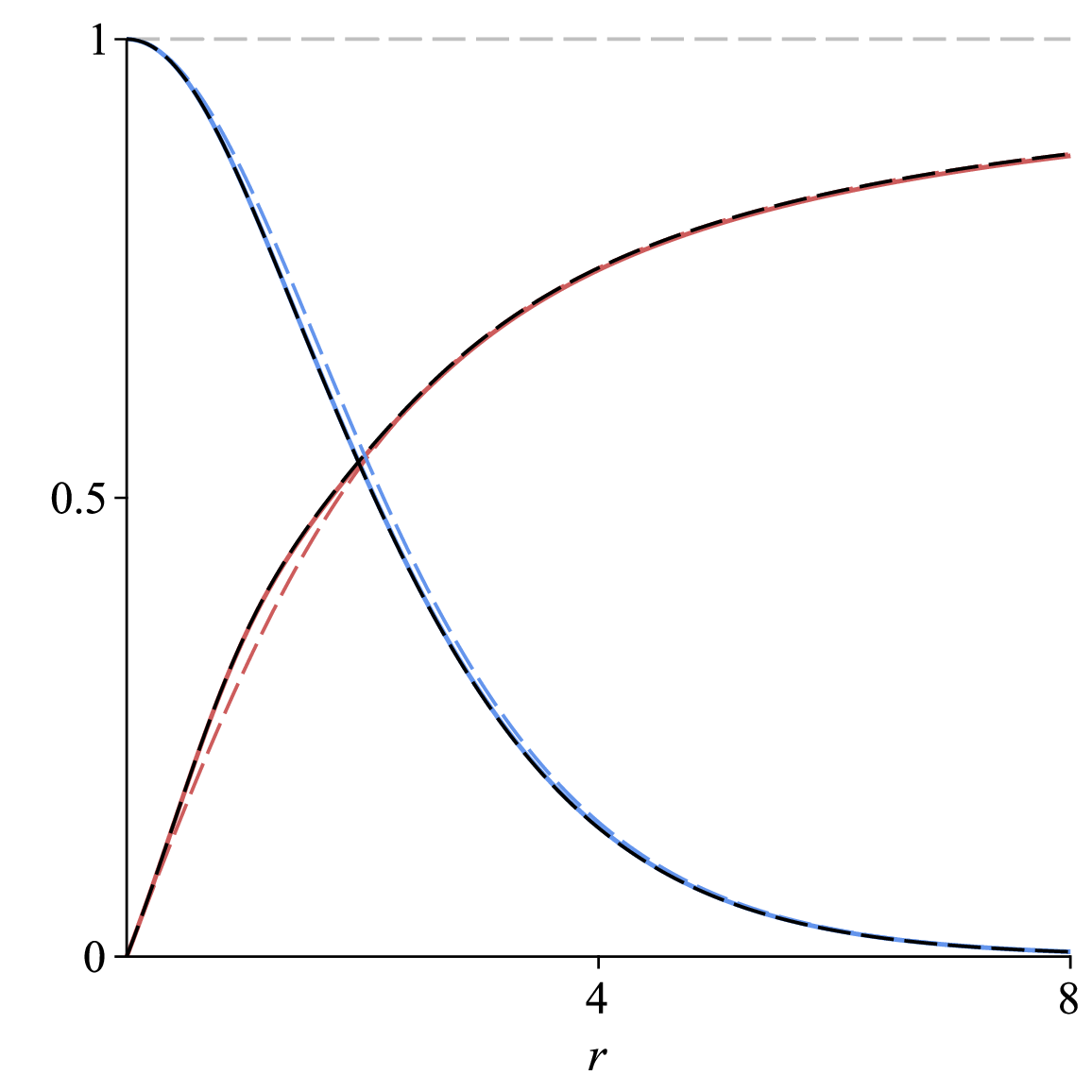}\includegraphics[width=0.33\linewidth]{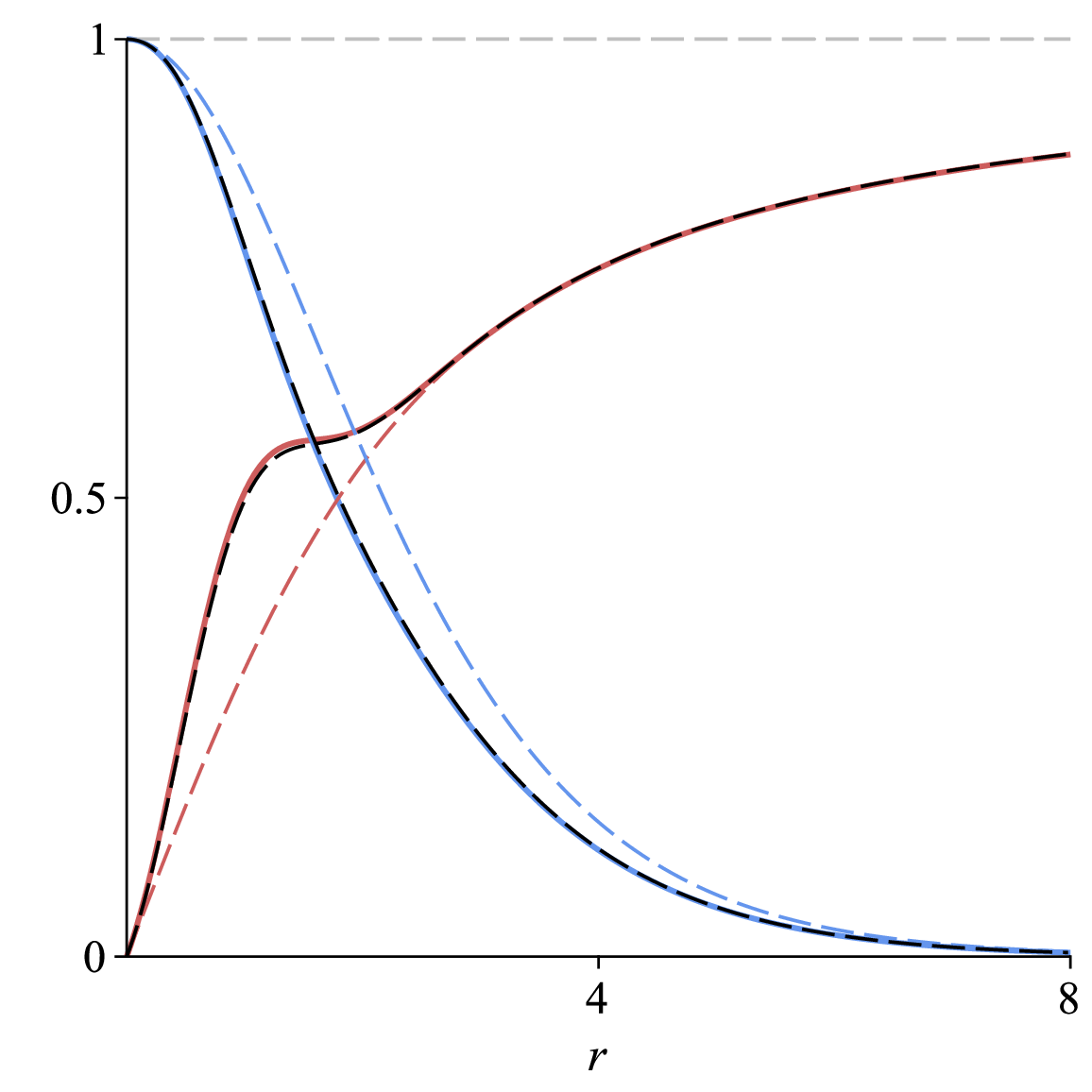}\includegraphics[width=0.33\linewidth]{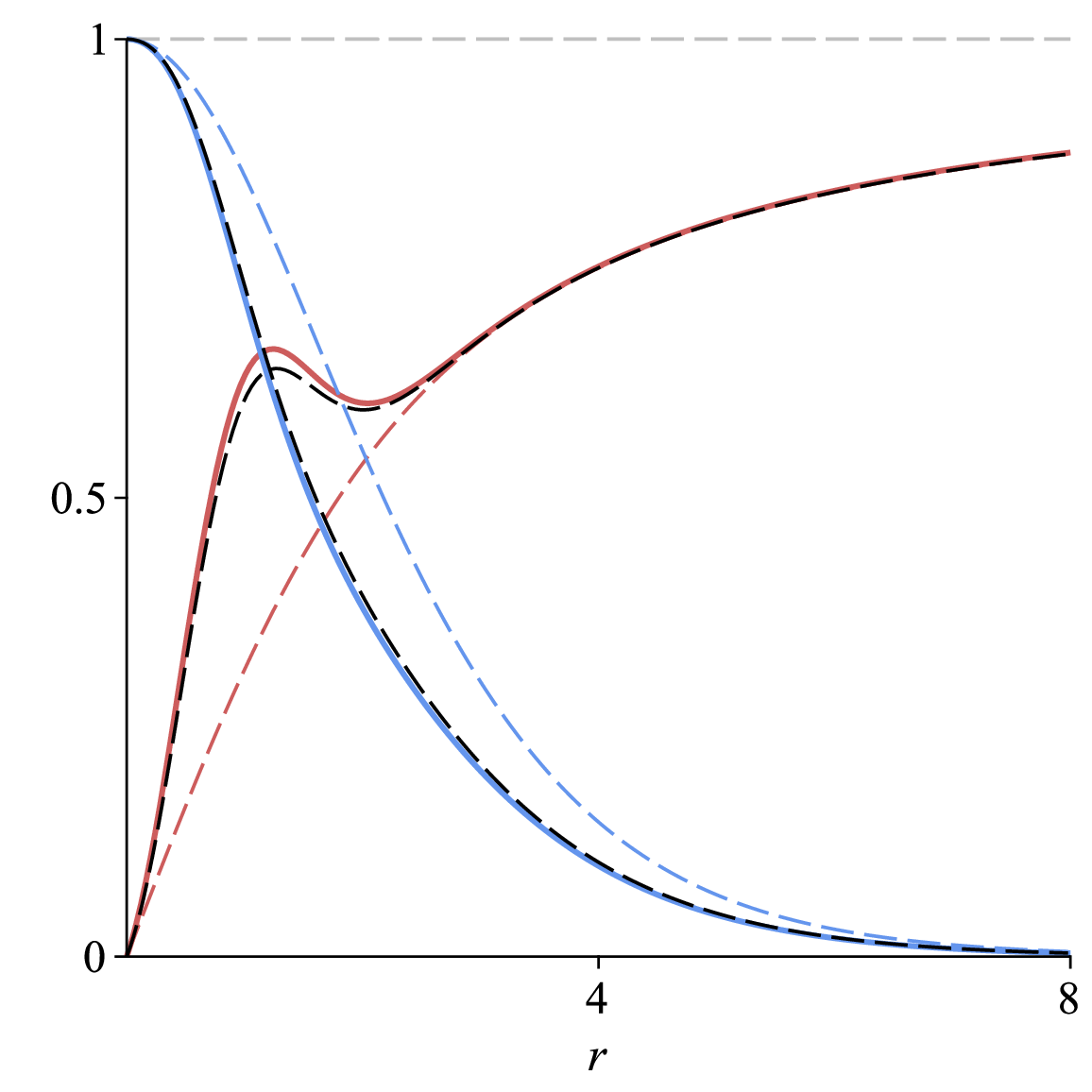}
	\caption{Solution $h(r)$ (red, solid line), $k(r)$ (blue, solid line) of equations~\eqref{FO} with $\alpha(r)$ and $\beta(r)$ given by~\eqref{Imp1}, with $\epsilon=0.1$ (left), $0.5$ (middle) and $0.8$ (right). Dashed lines of these same colors represent the Prasad-Sommerfield solution, while the black dashed lines correspond to the approximate solution~\eqref{approx}.}
	\label{ex1}
\end{figure}

Another class of models in which the asymptotic behavior of the impurity-free model is preserved is obtained from the choice
\begin{subequations}\label{Imp1}
	\begin{align}
		\alpha(r)&=\epsilon\left[ \frac{ \sqrt{\pi} }{2r^2}\csch^2(r)\text{erf}(r) + \frac{2e^{-r^2}}{r} \left(r^2 - \frac{\csch^2(r)}{2} - 1\right) \right],
	\\
	\beta(r)&=0.
\end{align}
\end{subequations}
This is the example we proposed in Subsection~\ref{reduc} as an illustration of the approximation~\eqref{37}, which we have used to deduce 
\begin{subequations}\label{approx}
\begin{align}
	h(r)&\approx \coth(r) -\frac{1}{r} +\epsilon r^2 e^{-r^2}, \\
	k(r)&\approx r\csch(r)e^{\epsilon\left(\frac{r}{2e^{r^2}}-\frac{\sqrt{\pi}\erf(r)}{4}\right)}.
\end{align}
\end{subequations}
These functions can be compared to the \qt{full} solution obtained by numerically solving the first-order equations. This has been done, and the result can be seen in Fig.~\ref{ex1}. One sees that, for $\epsilon=0.1$, which represents an impurity with amplitude of order $10^{-2}$, the solution is very similar to the PS result, although a deviation from the impurity-free case is discernible in the first half of the plot. The approximate solution~\eqref{approx} is however very close to the numerical result, with a difference too small to be seen in the plot. Numerical calculations show that a small difference exists between these functions, with a magnitude that changes with $r$, being greater at larger values. For $h(r)$, the approximation is roughly $0.002$ at its maximum, being of order $10^{-4}$ at $r=1$. At a neighborhood of the origin, these solutions are nearly identical, as the effect of $\alpha$ on the solution is of order $\epsilon r^2$. 

To allow for comparison, we have also solved these equations for two larger values of $\epsilon$, namely $\epsilon=0.5$ and $\epsilon=0.8$, with solutions depicted respectively in the middle and right panels of Fig.~\ref{ex1}. We see that, although the difference between the approximate and full solutions becomes gradually greater with greater $\epsilon$, it is still very good for $\epsilon=0.5$. The approximation does in fact appear quite reasonable even when $\epsilon=0.8$, although, in this case, the difference becomes visually distinguishable in the first half of the plot.

\subsection{Long-range solutions}
\begin{figure}[t]
 	\centering
 	\includegraphics[width=0.5\linewidth]{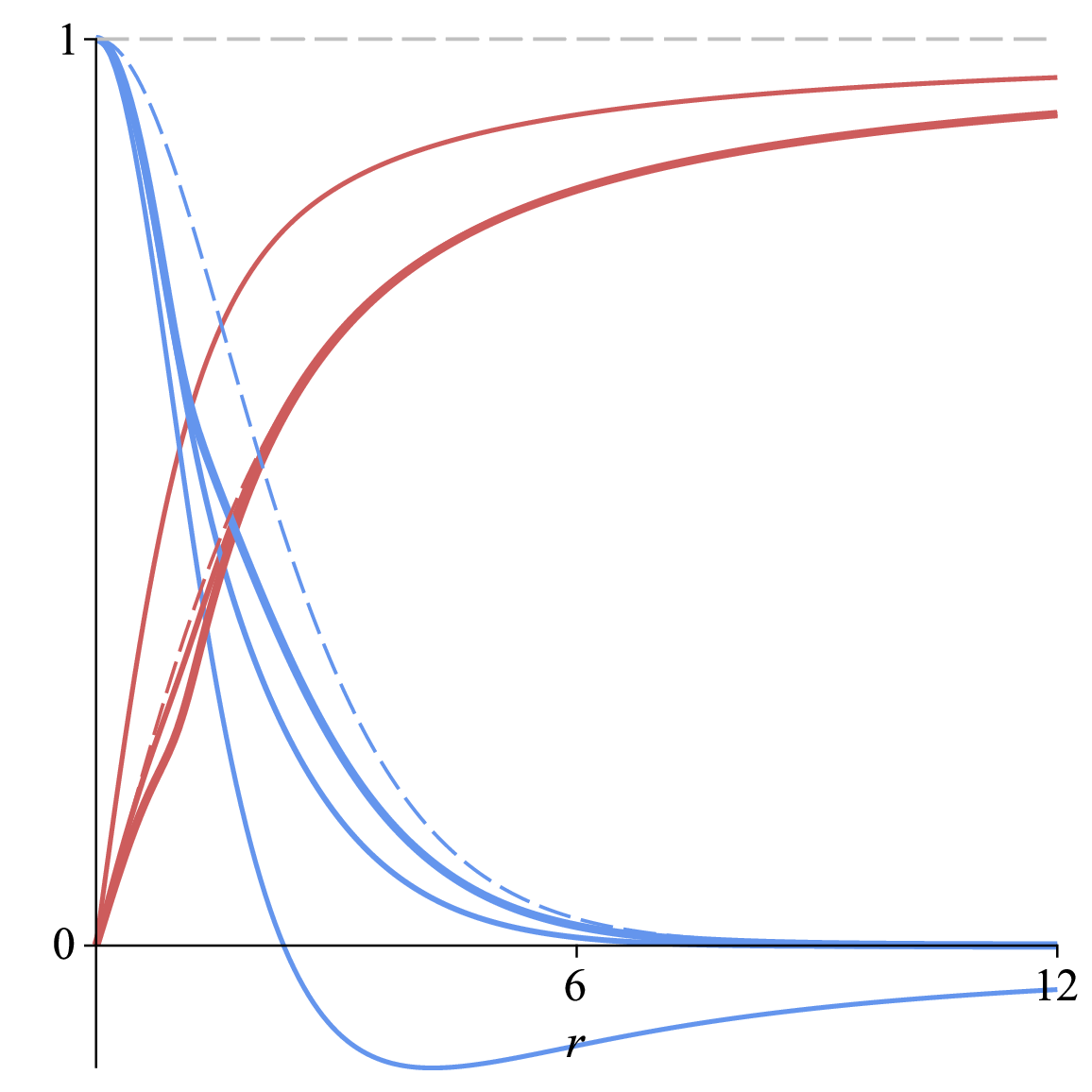}
 	\caption{Solution $h(r)$ (red, solid line), $k(r)$ (blue, solid line) of equations~\eqref{FO} with $\alpha=0$ and $\beta=\left(1-e^{-r^n}\right)/2r^n$. Line thickness increases with $n$, which takes the values $2$, $4$ and $8$, and dashed lines correspond to the Prasad-Sommerfield solution.}
 	\label{PL1}
 \end{figure}

Next, let us consider impurities leading to long-range monopoles, meaning that the monopole core will be enlarged due to impurity-induced changes in the asymptotic behavior of $k(r)$. One way this can be achieved is through the choice  
\begin{subequations}~\label{longrangeimp}
\begin{align}
	\alpha(r)&=Ae^{-r^2},\\ \beta(r)&=\frac{\left(1-e^{-r^n}\right)}{2r^n},
\end{align}
\end{subequations}
where $A$ is a real constant, with $n$ being a natural number different from one. As per the results of the asymptotic analysis conducted above, these impurities lead to fields that fall off to their asymptotic values according to a power law, regardless of the choice of $A$. Indeed, the choice $A=0$, amounting to a system of impurities completely described by the single function $\beta(r)$, is sufficient for this purpose.  This situation is depicted in Fig.~\ref{PL1} for three choices of $n$, namely $n=2$, $n=4$ and $n=8$. The two latter powers, and indeed any choice of with the exception of $n=2$ gives rise to a scalar field with the same asymptotic behavior as the Prasad-Sommerfield monopole, as the effect of these impurities amount to higher-order contributions at large $r$. The case $n=2$, however,  is different; the field still falls to unit as a power of $1/r$, but the coefficient is changed due to the $\beta(r)$ contribution, resulting in an asymptotic behavior of the form $1-1/2r$, as seen in Fig.~\ref{PL1}. On the other hand, the large $r$ properties of $k(r)$ are always changed, since any power of $1/r$ is faster than the exponential behavior of the PS solution. Indeed, the large $r$ behavior of $k(r)$ is of the form $\kappa(r)\approx -r^{1-n}/2$, agreeing with the results found in subsection~\ref{asympt}. This behavior is shown in Fig.~\ref{PL1} for three values of n. Although the scale of the plot only allows the graphic visualization of this phenomenon for the $n=2$ case, it can be verified that the range of the monopole decreases progressively with $n$, which physically amounts to increasingly larger monopole cores. The $n=4$ and $n=8$ cases may at first appear to contradict our results, as the former approaches the $r$ axis faster than the latter. However, it must be remembered that $k(r)\sim -1/r^{n-1}$, being thus ultimately increasing at large $r$, for any choice of $n\neq 0$. Thus, solutions must first cross the $r$-axis and then increase after reaching a minimum. With an appropriate close-up, it can be seen that both $n=4$ and $n=8$ do indeed display the predicted behavior and approach zero as $-1/r^{3}$ and $-1/r^7$ respectively, although the scale of the plot and finite line thickness prevent us from graphically representing these behaviors simultaneously.

\begin{figure}[h]
    \centering
 	\includegraphics[width=0.5\linewidth]{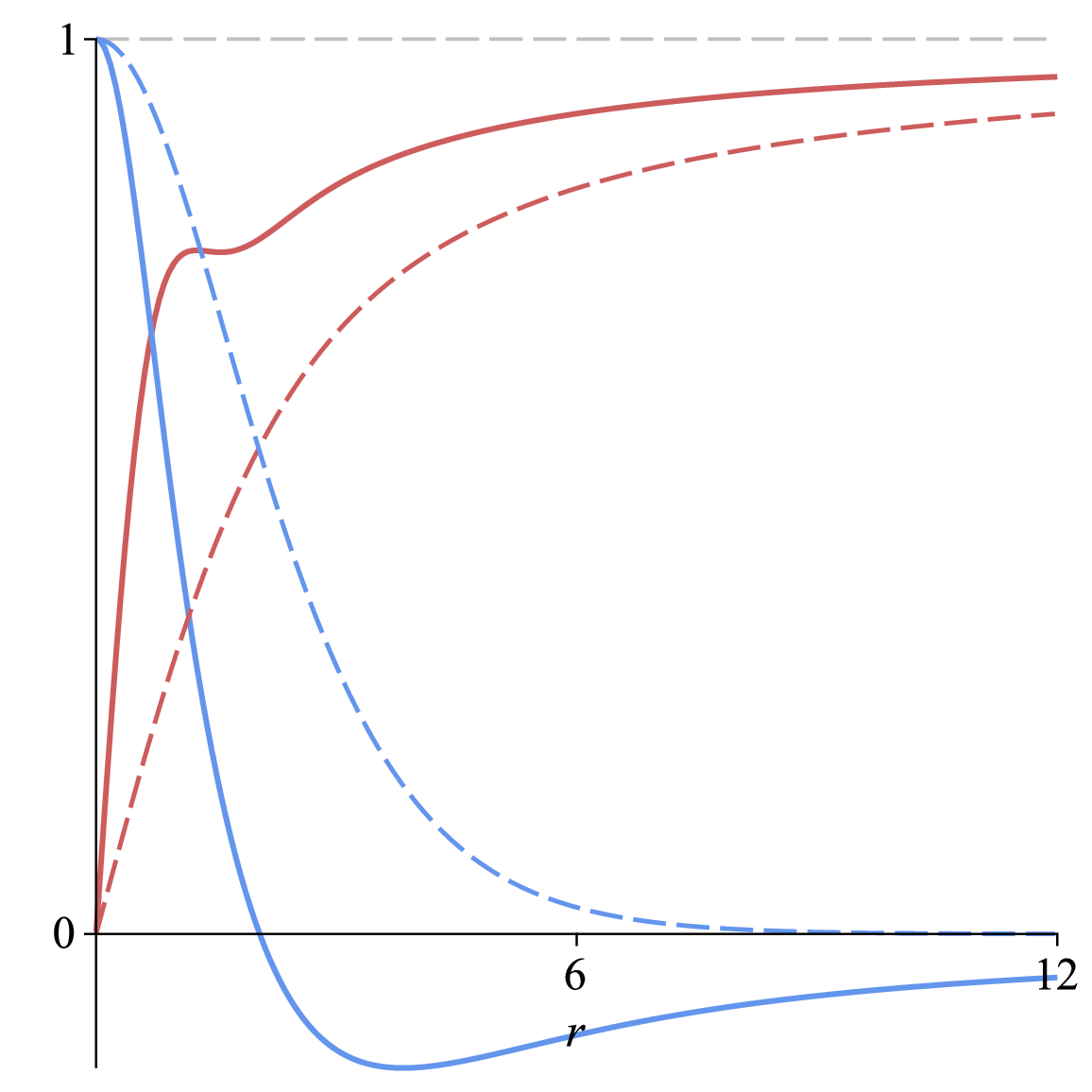}\includegraphics[width=0.5\linewidth]{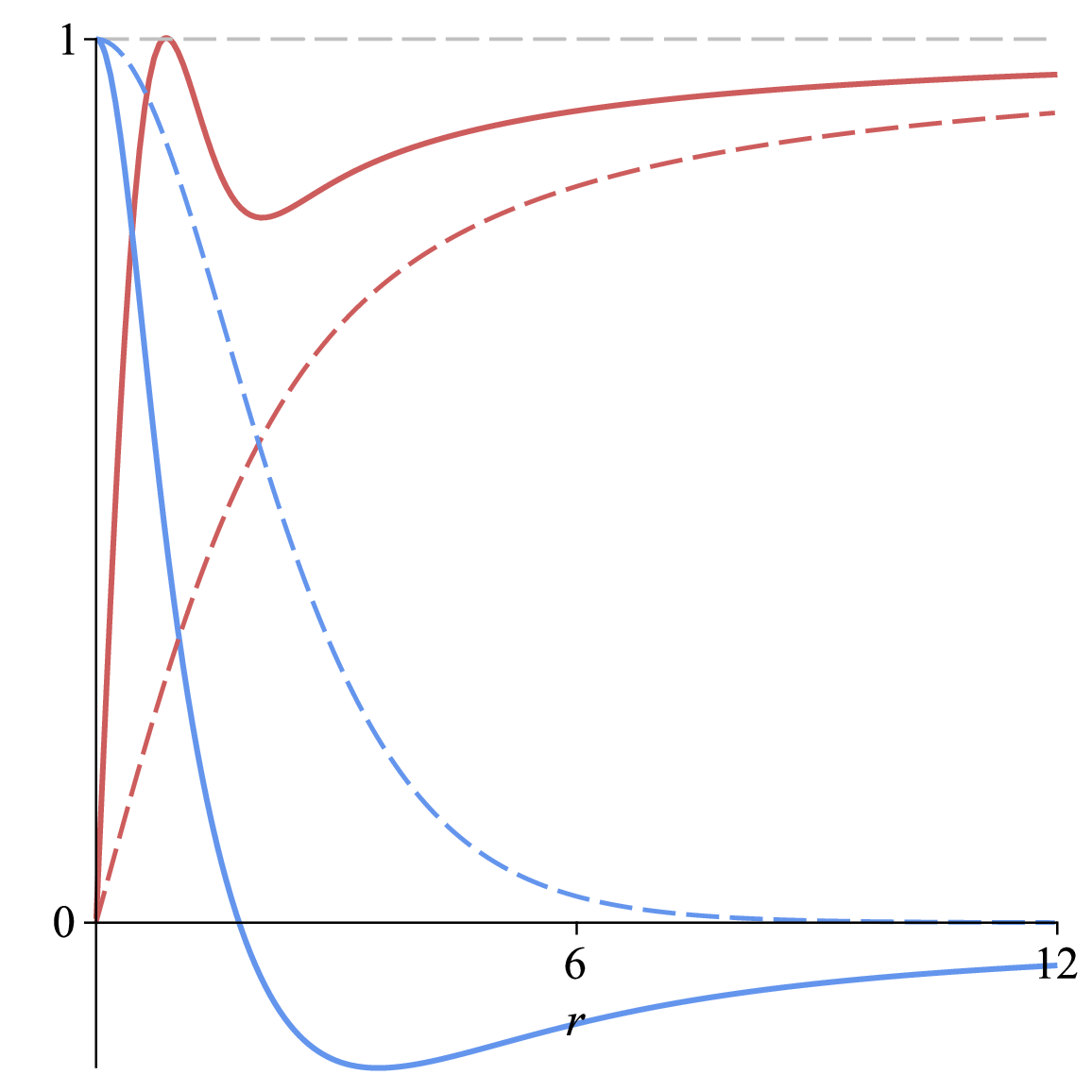}
 	\caption{Solution $h(r)$ (red, solid line), $k(r)$ (blue, solid line) of equations~\eqref{FO} with $\alpha(r)$ and $\beta(r)$ given by~\eqref{longrangeimp}, with $A=1$ (left) $A=2$ (right) and $n=2$. Dashed lines represent the Prasad-Sommerfield solution.}
 	\label{PL2}
\end{figure}

Although the asymptotic behavior of the solution is independent of $A$, the exponential function plays a large role in the interior of the monopole, where $\alpha$ engenders a complex internal structure which is highly dependent on the choice of $A$, in a manner similar to what we have found in our previous examples. This pattern is exemplified in Fig.~\ref{PL2} for the case $n=2$. Together with the $A=0$ example depicted in Fig.~\ref{PL1}, these solutions represent three members of the family of impurities specified by~\eqref{longrangeimp}, differing by the choice of $A$. We note the emergence of maxima and minima located at distances of order one from the origin. Within a sufficiently small neighborhood of the origin, however, the fields present the same behavior as the standard PS monopole, with $h(r)\approx C_1r$, $k(r)\approx C_1r^2/2$.

\subsection{Super long-range solutions}
\begin{figure}[b]
 	\centering
\includegraphics[width=0.5\linewidth]{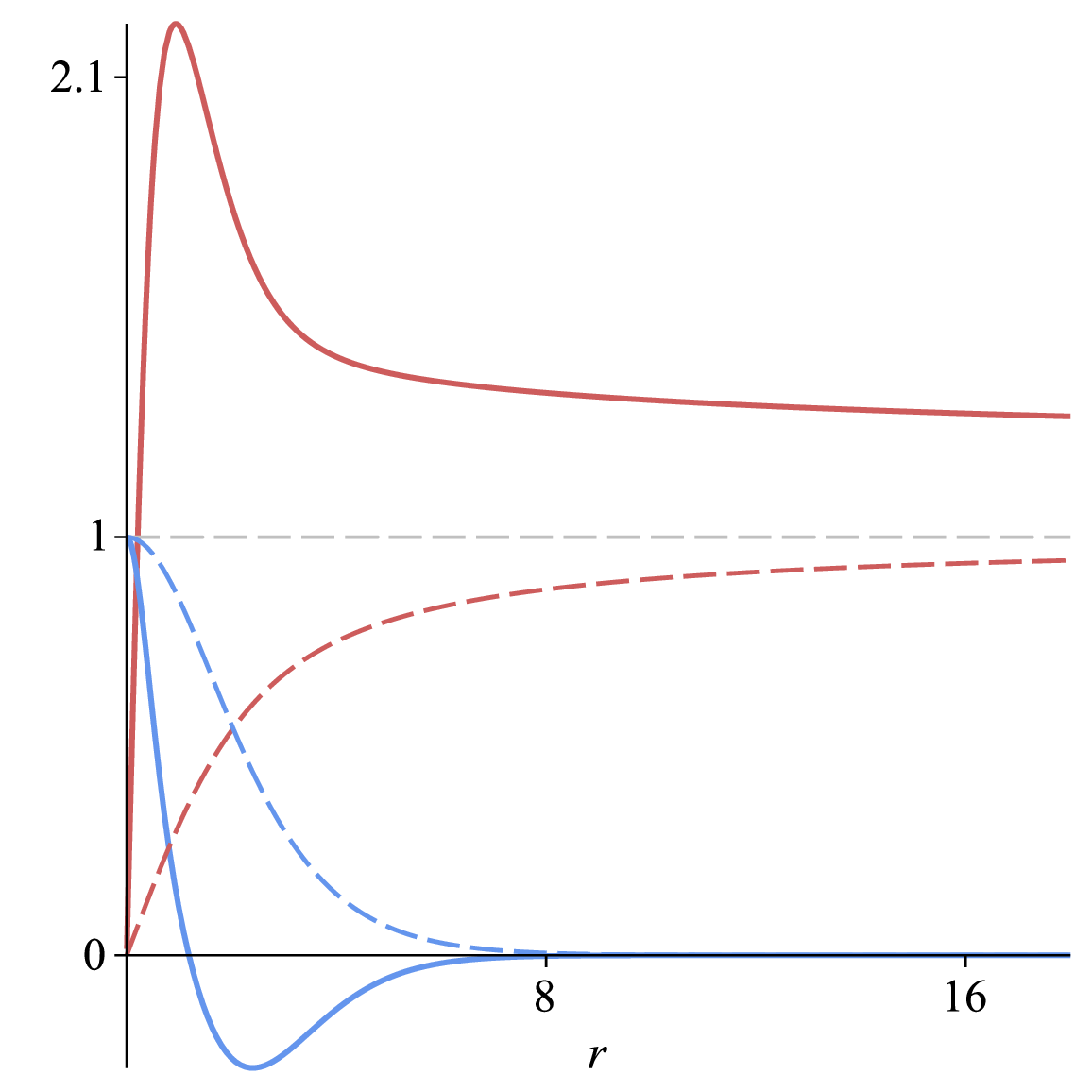}\includegraphics[width=0.5\linewidth]{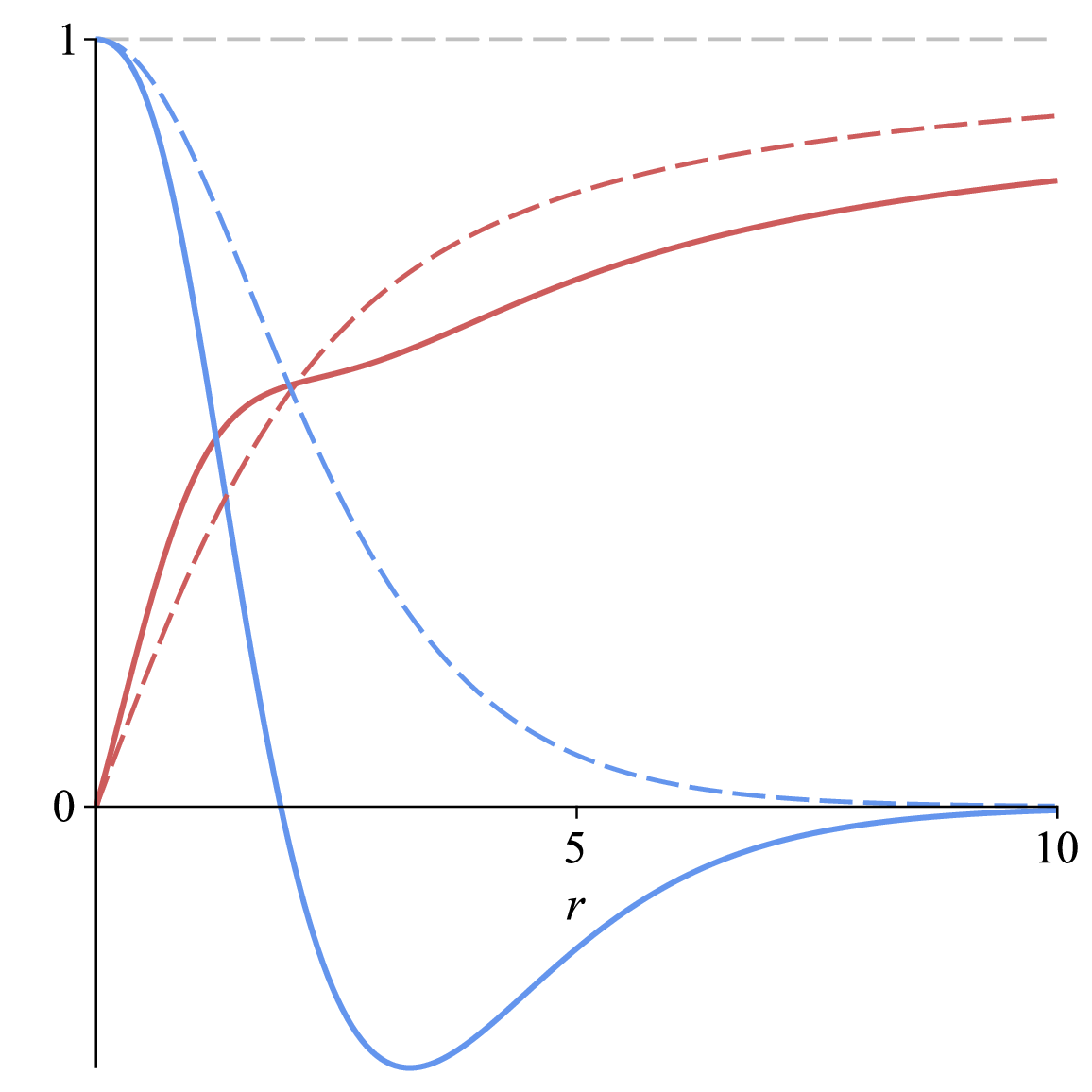}
 	\caption{Solution $h(r)$ (red, solid line), $k(r)$ (blue, solid line) of equations~\eqref{FO} with $\alpha(r)$ and $\beta(r)$ given by~\eqref{impurityex2}, with $A=1$ (left) and $A=-0.2$ (right). Dashed lines represent the Prasad-Sommerfield solution.}
 	\label{ex2fields}
 \end{figure}

Now, let us exemplify the case, discussed above, of logarithmically-decaying scalar fields. To this end, we use an impurity specified by the functions
\begin{subequations}\label{impurityex2}
 	\begin{align}
 		\alpha(r) &=\frac{A\tanh^4(r)}{r^3\ln^2(r+1)}, \label{impurityex2a}\\ 
 		\beta(r) &=r^2\sech^2(r),\label{impurityex2b}
 	\end{align} 
 \end{subequations}
where $A$ is a real constant. At large $r$, these functions give rise to a function $f(r)$ of the form~\eqref{implog}, generating an $h(r)$ with super long-range behavior~\eqref{hlog}, while $k(r)$ goes to zero exponentially fast.  We have solved the first-order equations for this model in the cases $A=1$ and $A=-0.2$, and the solutions are depicted respectively in Fig.~\ref{ex2fields}. In both of these cases, $h(r)$ presents a much longer range in comparison to the PS solution, with the sign of $A$ determining whether the $h=1$ asymptote is reached from above or below. The rate at which $k(r)$ approaches zero is also affected by this constant, and we see in particular that, for $A=-0.2$, the monopole core is enlarged, as $k(r)$ reaches zero after a greater distance.
\cmmnt{
 \begin{subequations}\label{impurityex2old}
 	\begin{align}
 		\alpha(r) &=\frac{\tanh^4(r)}{r^2}\left(\frac{1}{r^2}+\frac{A}{r\ln^2(r)+1}\right), \label{impurityex2aold} \\
 		\beta(r) &=Br^2\sech^2(r),\label{impurityex2bold}
 	\end{align} 
 \end{subequations} }
Since the logarithmic dependence in this example is generated solely by $\alpha(r)$, the monopole core size is of the same order as in the impurity-free case, while $h(r)$ remains nontrivial long after the scalar field has left the monopole core. Although we have not included an explicit example of this kind, it is also possible to obtain a function $k(r)$ which goes to zero logarithmically. This can be achieved, for example, with an impurity such that $\beta(r)\sim C\left[  \left(r\ln(r) -\left(r\ln(r) \right)^{-1}\right)^{-2}\right]$, which upon integration gives $k(r)\sim C/\ln(r)$ at large $r$.

\section{Remarks about dyons}\label{dyons}

Although we have thus far focused on pure magnetic monopoles, the presence of impurities does not preclude defects with electric charge. In the case of topologically nontrivial configurations, this possibility implies localized structures with both magnetic and electric charge, called dyons. In the standard theory, such solutions have first been shown to exist by Julia and Zee~\cite{juliaandzee}, an analysis which is greatly simplified by the formal analogy between the field equations for $A_{\0}$ and $\phi$ in the Hedgehog gauge. As we shall see, this analogy is however not preserved when impurities are added to the system.

The BPS bound for dyons is of a different nature than that of pure monopoles, as it depends on a Noether (electric) charge as well as the topological one. In the supersymmetric theory, this situation signals the conservation of two Supercharges, as shown by Witten~\cite{PLBWitten}. This generalization may however be easily dealt with through recourse to the strategy conveyed in~\cite{dentadyon}, which still works for the models considered in this paper. To this end, we introduce real constants $\mu$ and $\nu$ such that 
\begin{equation}
	\mu^2+\nu^2=1,
\end{equation}
which, in the $\lambda\to 0$ limit, we use to rewrite the energy functional in the form
 \begin{equation}
	\begin{aligned}
		E &=\frac{1}{2}\int_{}d^3x|D_{\0}\phi|^2-\int_{} d^3x\left\{ \mathrm{Tr}\left[(E_k + \nu D_k\phi)^2 + (B_k + \mu D_k\phi + S_k)^2\right]\right\}+4\pi \mu N + |q_e|\nu\\
		&  \geq 4\pi \mu N + |q_e|\nu 
	\end{aligned}
\end{equation}
where $q_e$ is the electric charge, derived from the integral version of Gauss's law $\int d^3x  \nabla\cdot\mathbf{e} =q_e$. The bound shown on the last line can only be saturated by solutions of the first-order equations
\begin{subequations}\label{BPSDyon}
	\begin{align}
		D_{\0}\phi&=0, \label{BPSD0}\\ 
		E_k &=-\nu D_k\phi, \label{BPSEk} \\
		B_k&=-\mu D_k\phi-S_k.	\label{BPSBk}
	\end{align}
\end{subequations}

\cmmnt{Substitution of~\eqref{BPSEk} into~\eqref{BPSBk} and calculation of the flux integral the result leads to $q_e=4\pi N\nu/\mu$, since localized impurities cannot contribute to the flux at infinity. We may thus write $\nu$ and $\mu$ in terms of these charges, so that the Bogomol'nyi energy for dyon solutions solving these equations can be written in the form
\begin{equation}\label{DBound}
	E= \sqrt{q_m^2 + q_e^2}.
\end{equation}}

If a dyon solution exists, use can be made of the first-order equations, together with the definition of electric and magnetic charges, to deduce a Bogomol'nyi energy $E= \sqrt{q_m^2 + q_e^2}$. However, it can be shown that equality is in general unattainable for arbitrary $S_k$. Indeed, Eqs.~\eqref{BPSDyon} imply
\begin{equation}
	\begin{split}
D_kE_k&=\frac{\nu}{\mu}D_k\left(B_k+S_k\right)\\
&=\frac{\nu}{\mu}D_kS_k,
	\end{split}
\end{equation}
where in the last line use was made of the Bianchi identity, which implies $D_kB_k=0$. However, the fields are still required to satisfy~\eqref{gauss} which, when combined with~\eqref{BPSD0}, leads to $D_kE_k=0$. Thus,~\eqref{BPSDyon} can be consistent with Gauss' law only if either $\nu=0$, which is the case of the monopoles dealt with in Subsection~\ref{BPSsec}, or if $D_kS_k=0$, which places a strong constraint in the impurity functions.

\cmmnt{
	
The spherically symmetric Bogomol'nyi equations in this case are 	
\begin{subequations}\label{FOD}
	\begin{align}
	&	\mu h'=\frac{1-k^2}{r^2} -\alpha r^2 -\beta, \label{FOAD}\\
	&	\frac{k'}{r}=-\left(\nu\frac{hk}{r}+\beta\right), \\\label{FOBD}
			&j(r)=\nu h(r).
	\end{align}
\end{subequations}	
Because $\phi$ is covariantly constant, solutions that are time-independent in appropriate gauges are possible if $[\phi, A_{\0}]=0$, which suggests gauge fields with a temporal component parallel to $\phi$. In the spherically symmetric case, we thus complement the ansatz with
\begin{equation}\label{juliazee}
	A^a_\0=\frac {j(r)}{r}x^a,
\end{equation}
and must now supplement the field equations the additional conditions 
\begin{align}
	\lim_{r\to\infty}j(r)=C + \frac{q_e}{4\pi r},   &&
	\lim_{r\to0}j(r) = 0	\label{J2},	
\end{align}
where $C$ is a real constant.
}

We could also further generalize our model by adding an electric coupling similar to the one we have considered for $B_k$, meaning an impurity Lagrangian of the form
\begin{equation}\label{Limpp}
\LL_{imp}=2\Tr{\left(\bar{S}_k+S_k\right)D_k\phi+S_kB_k +\bar S_kE_k}.
	\end{equation}
In that case, Eq.~\eqref{BPSEk} would be generalized to
\begin{equation}
		E_k =-\nu\left( D_k\phi -\bar S_k\right),
\end{equation}
while the remaining BPS equations remain the same as in~\eqref{BPSDyon}. This would lead to a theory that is in many respects similar to what has been discussed within this paper, although with the important difference that truly static configurations would not be allowed, as no configuration with $E_k=0$ could solve Gauss's law. The generalization of  our results to this case is straightforward due to the analogy between the old and new couplings, although saturation of the Bogomol'nyi bound and consistency with Gauss' Law would still require constraining the impurity functions.

Finally, one could follow~\cite{WilsonLines} and add an impurity coupled directly to $A_{\0}$, thus adding a new source term to Gauss's law, while leaving Eqs.~\eqref{BPS} unchanged. In this case, stationary BPS solutions can be solved in the exact same way, with the result fed into Gauss's Law to determine $A_{\0}$. This has important consequences for monopole scattering as the  kinetic energy of the model is changed in a nontrivial way. The basic principles of this method, which should still be applicable to the models considered in this work, have been thoroughly explored in~\cite{WilsonLines}.

\section{Conclusion and perspectives}\label{end}

In this work, the well-known Yang-Mills-Higgs theory has been generalized to include couplings with localized impurities. We have first considered scalar impurities, which have been shown to play the role of an effective permeability in the Yang-Mills equations, resulting in models which, in the Bogomol'nyi limit, allow for solutions formally identical to some of the systems previously considered in Refs.~\cite{internal, bimag, multimag}. Following the approach of~\cite{Hook}, we have introduced a novel class of models governed by Lagrangians of the form~\eqref{laggen}, with impurity coupling specified by~\eqref{Limp}. The form of this Lagrangian has been chosen in such a way as to preserve half the BPS sectors of the original model, namely the monopole sectors. The resulting Bogomol'nyi bound is precisely the same found in the $N>0$ sectors of impurity-free theory, while the associated BPS equations are each deformed by a localized, position-dependent function $S_k(\mathbf{x})$, representing a fixed $su(2)$ impurity coupled to the scalar and gauge fields of the model. These impurities represent external fields that may be associated to background fields transforming under the adjoint representation of $SU(2)$. We have investigated some of the theoretical features of this model, with emphasis on the energy minimizers found in the Bogomol'nyi limit, which are valuable analytical tools due to their relative simplicity and ensured stability. We have seen that the localized character of these impurities ensures preservation of the most important topological features of this model, including the emergence of a magnetic charge for every finite energy solution in a topologically nontrivial sector. 


If the impurities are spherically symmetric, the well-known hedgehog ansatz can be used to reduce the problem to a system of two differential equations. We have used impurities with this symmetry to derive a spherically symmetric energy functional, from which field equations preserving this symmetry can be derived. A detailed analysis of the symmetrical BPS equations has been conducted, including investigation of the changes caused by impurities on the asymptotic properties of solutions. It is shown that the long-distance behavior of the monopole can be significantly changed due to impurity coupling. This fact is important to monopole dynamics, which is itself a relevant problem in modern physics, as monopole scattering plays a large role in attempts to solve the monopole problem, which include the constraining of inflationary models. A reduction of the system of first-order equations to equivalent integro-differential or second-order ODEs for a single function has been achieved in many cases, and an approximation valid for small-amplitude impurities has been obtained for some models, with satisfactory numerical results for the example considered. Finally, we have chosen specific impurities to present concrete examples that illustrate some important features of the theory, obtaining solutions with a richer internal structure within the monopole core, and modified asymptotic behavior at large distances from it.

Several extensions of our results are possible. One notable application that has been met with significant success in recent years is the investigation of defect scattering in the presence of impurities, a path which has led to the discovery of interesting phenomena in other impurity-defect settings. To this end, the asymptotic analysis developed here may be significant, as those should play a large role in the long-distance interactions of multi-monopole systems. Another perspective lies in the investigations of multi-monopole solutions, for which the ansatz~\eqref{hedgehog} is not applicable. Historically, that was first achieved for $N=2$ using an axially symmetric ansatz~\cite{PRL45}. That approach would however be of limited applicability in the present scenario, as it would require that both the monopoles and impurities lie along the same axis. Other successful approaches do however exist, such as the iterative generation of multi-monopole solutions through Bäcklund transformations~\cite{PLB99} and the Nahm approach, which has been very successful in investigations of this kind~\cite{NahmI, NahmII, NahmIII}, although generalization of these results to the presently considered case is not a trivial matter.

Another path for possible generalization is found in models with electric impurities, such as those considered in~\cite{WilsonLines} or those obtained from a Lagrangian of the form~\eqref{Limpp}. The introduction of electric fields gives rise to a kinetic energy in the model, which may generate interesting effects in combination with the impurity models presented here. 
Another possibility concerns the study of magnetic monopoles in the presence of impurities, in the case where the localized solutions engender multimagnetic structures, which arise from an appropriate extension of the SU(2) gauge group \cite{multimag}.
We can also investigate collisions involving magnetic monopoles in the presence of impurities. A similar problem has been numerically investigated, in the impurity-free scenario, by several authors, see for example \cite{Forgacs, MantonSamols,CMM}.  This line of investigation may also  benefit from previous studies on the scattering of vortices with impurities in the plane; see, for instance, \cite{Cockburn,Kruch,BCM} and references therein.

\acknowledgments{This work is supported by the Brazilian agency Conselho Nacional de Desenvolvimento Cient\'ifico e Tecnol\'ogico (CNPq), grants Nos. 402830/2023-7 (DB and MAM), 303469/2019-6 (DB), 151204/2024-1 (MAL) and 306151/2022-7 (MAM).}



\end{document}